\documentclass{article}
\usepackage[utf8]{inputenc}
\usepackage{mathtools}
\usepackage{amssymb}
\usepackage{amsthm}
\usepackage{amsmath}
\usepackage{booktabs}
\usepackage{geometry}
\usepackage{authblk}
\usepackage{hyperref}

\usepackage{comment}

\newtheorem{theorem}{Theorem}
\newtheorem{corollary}{Corollary}
\newtheorem{conjecture}{Conjecture}
\newtheorem{lemma}{Lemma}
\newtheorem{definition}{Definition}
\theoremstyle{remark}
\newtheorem{remark}{Remark}

\usepackage{xcolor}

\newcommand{\F}{\mathbb{F}}

\newcommand\trace{\mathrm{Tr}}
\newcommand\LS{\mathrm{LS}}
\newcommand\Lin{\mathrm{Lin}}
\newcommand\sylvester{\mathrm{Syl}}
\newcommand\resultant{\mathrm{Res}}
\newcommand\im{\mathrm{Im}}
\newcommand\wt{\mathrm{wt}}

\newcommand\scalarprodV[2]{\langle #1, #2 \rangle_{\mathcal{V}}}

\newcommand\walshC[1]{\widehat{ #1 }}

\renewcommand\footnotemark{}
\title{A Further Study of Quadratic APN Permutations in Dimension Nine\thanks{This  work  was  funded  by Deutsche  Forschungsgemeinschaft  (DFG);  project  number 411879806. The research of C. Carlet is partly supported by the Trond Mohn Foundation and Norwegian Research Council.

This manuscript version has been accepted for publication, after peer review
but is not the Version of Record and does not reflect post-acceptance improvements, or any
corrections. The Version of Record is available online at: \url{https://doi.org/10.1016/j.ffa.2022.102049}.}}

\date{} 

\author[1]{Christof Beierle}
\affil[1]{Ruhr University Bochum, Universit\"atsstra\ss e 150, 44801 Bochum, Germany}
\author[2,3]{Claude Carlet}
\affil[2]{LAGA, University of Paris 8, Saint-Denis, France}
\affil[3]{Department of Informatics, University of Bergen, PB 7803, 5020 Bergen, Norway}
\author[1]{Gregor Leander}
\author[4]{L\'eo Perrin}
\affil[4]{Inria, 2 rue Simone Iff, 75012, Paris, France}
\begin{document}
\maketitle
\begin{abstract}
    Recently, Beierle and Leander found two new sporadic quadratic APN permutations in dimension 9. Up to EA-equivalence, we present a single trivariate representation of those two permutations as $C_u \colon (\F_{2^m})^3 \rightarrow (\F_{2^m})^3, (x,y,z) \mapsto (x^3+uy^2z, y^3+uxz^2,z^3+ux^2y)$, where $m=3$ and $u \in \F_{2^3}\setminus\{0,1\}$ such that the two permutations correspond to different choices of $u$.  We then analyze the differential uniformity and the nonlinearity of $C_u$ in a more general case. In particular, for $m \geq 3$ being a multiple of 3 and $u \in \F_{2^m}$ not being a 7-th power, we show that the differential uniformity of $C_u$ is bounded above by 8, and that the linearity of $C_u$ is bounded above by $8^{1+\lfloor \frac{m}{2} \rfloor}$. Based on numerical experiments, we conjecture that $C_u$ is not APN if $m$ is greater than $3$. We also analyze the CCZ-equivalence classes of the quadratic APN permutations in dimension 9 known so far and derive a lower bound on the number of their EA-equivalence classes. We further show that the two sporadic APN permutations share an interesting similarity with Gold APN permutations in odd dimension divisible by 3, namely that a permutation EA-inequivalent to those sporadic APN permutations and their inverses can be obtained by just applying EA transformations and inversion to the original permutations. 
    
\end{abstract}

\section{Introduction}
Almost perfect non-linear (APN) functions are of great interest, both from a theoretical perspective as well as from a practical point of view as those functions provide the (locally) optimal resistance against differential attacks~\cite{DBLP:journals/joc/BihamS91} on symmetric cryptographic primitives. For being of interest in the design of symmetric primitives, in particular in substitution-permutation networks for  block ciphers and cryptographic permutations, those functions need to be permutations themselves. However, our  understanding of APN functions in general and of APN permutations in particular is rather limited.

Indeed, at the time of writing, all of the APN permutations known so far belong (up to CCZ-equivalence) to one of the four cases listed below. The first two cases define infinite families of functions, while the last two cases are sporadic instances of APN permutations and have not been classified into infinite families of APN functions yet.

\begin{enumerate}
    \item APN monomial functions $F\colon \F_{2^n} \rightarrow \F_{2^n}$ for $n$ being odd.
    \item The quadratic functions $F\colon \F_{2^{3k}} \mapsto \F_{2^{3k}}, x \mapsto x^{2^s+1} + \omega x^{2^{ik}+2^{tk+s}}$, where $s,k$ are positive integers with $k$ being odd, $\gcd(k,3)=\gcd(s,3k)=1$, $i=sk \mod 3$, $t=3-i$, and $\omega \in \F_{2^{3k}}^*$ with order $2^{2k}+2^k+1$ (Corollary 1 of~\cite{DBLP:journals/tit/BudaghyanCL08}).
    \item Dillon's permutation~\cite{Dillon-perm}, i.e., a permutation CCZ-equivalent to the (non-bijective) Kim function $F\colon \F_{2^6} \mapsto \F_{2^6}, x \mapsto x^3 + \alpha x^{24} + x^{10}$, where $\alpha \in \F_{2^6}^*$ is an element with minimal polynomial $X^6 + X^4 + X^3 + X + 1 \in \F_2[X]$.
    \item The two quadratic APN permutations
    \begin{align*} F_0 \colon \F_{2^9} \rightarrow \F_{2^9}, \quad &x \mapsto  x^3 + u^2x^{10} + ux^{24} + u^4x^{80}+ u^6x^{136} \\
F_1 \colon \F_{2^9} \rightarrow \F_{2^9}, \quad &x \mapsto x^3 + ux^{10} + u^2x^{17} + u^4x^{80} + u^5x^{192}\;,
\end{align*}
where $u \in \F_{2^9}$ is a zero of $X^3+X+1$ (\cite{DBLP:journals/corr/abs-2009-07204}).
\end{enumerate}

One of the most interesting problems in this setting is to find infinite families of APN functions that cover the sporadic examples. This would on the one hand give -- hopefully -- more APN functions or even APN permutations, and on the other hand, not less important, provide further insights into the topic.

Dillon's permutation was found in~\cite{DBLP:conf/crypto/PerrinUB16} to be a particular case of a specific structure called the \emph{butterfly}. More precisely, let $R_y(x) = (x + \alpha y)^3 + y^3$ be a bivariate function of $\F_{2^3}$ and $\alpha$ be a non-zero element of $\F_{2^3}$ with a trace equal to 0, so that $R_y$ is always a permutation. Then Dillon's permutation is affine-equivalent to $(x,y) \mapsto \left(R_y^{-1}(x), R_{R_y^{-1}(x)}(y) \right)$. Unlike Dillon's original permutation, this bivariate structure can easily be defined using larger fields $\F_{2^n}$ as long as $n$ is odd. It was also shown to always be at most differentially 4-uniform in~\cite{DBLP:conf/crypto/PerrinUB16}, and follow-up works showed that further generalizations had the same property along with the best known non-linearity~\cite{DBLP:journals/tit/CanteautDP17,Fu_Feng_Wu_2017,Li_Tian_Yu_Wang_2018}. Being an infinite family containing Dillon's permutation, there was of course hope that it might yield a solution to the big APN problem, but unfortunately it was later proved that it is impossible for a generalized butterfly to be APN unless it operates on 6 bits~\cite{DBLP:journals/ccds/CanteautPT19}.

The two APN permutations $F_0$ and $F_1$ have been found only very recently, and as such are not known to be part of any larger families of APN functions. The authors of~\cite{DBLP:journals/corr/abs-2009-07204} found
more than 12,000 new instances of quadratic APN functions in dimension $n=8$, as well as 35, resp., 5 new instances of quadratic APN functions in dimension $n=9$, resp., $n=10$. Out of all those APN functions found in~\cite{DBLP:journals/corr/abs-2009-07204}, only two are CCZ-equivalent to permutations, i.e., $F_0$ and $F_1$ as defined above. Note that many instances of quadratic APN functions have been provided in the literature before, see, e.g., \cite{browning2009apn,DBLP:journals/amco/EdelP09,weng2013quadratic,DBLP:journals/dcc/YuWL14}.

\subsection{Our Contribution and Results}
In this paper, we study the recently-found 9-bit APN permutations $F_0$ and $F_1$. In the first part,
we observe that they both have an EA-equivalent representation as a quadratic rotation-symmetric trivariate function \[C_u \colon \F_{2^m} \times \F_{2^m} \times \F_{2^m} \rightarrow \F_{2^m} \times \F_{2^m} \times \F_{2^m}, \quad (x,y,z) \mapsto (x^3+uy^2z, y^3+uxz^2,z^3+ux^2y)\]
for $m=3$ and $u \in \F_{2^3} \setminus \{0,1\}$. With those parameters for $m$ and $u$, the function $C_u$ is also a permutation. We outline our approach to obtain this trivariate representation by utilizing a generalization of the TU-decomposition~\cite{DBLP:conf/eurocrypt/BiryukovPU16}. We then show that the two sporadic APN permutations in dimension 9 are contained in an infinite family of differentially $d$-uniform functions with $d\leq 8$. In particular, for $m$ being a multiple of 3, we show that the differential uniformity of $C_u$ is bounded above by 8 if and only if $u \in \F_{2^m} \setminus \{0\}$ is not a 7-th power. We further study the linearity of $C_u$ with $m$ being a multiple of 3 and $u \in \F_{2^m}\setminus\{0\}$ not being a 7-th power and show that it is bounded above by $8^{1+\lfloor m/2 \rfloor}$.
Thus, if $m$ is odd, $\gcd(m,3)=3$, and $u \in \F_{2^m}$ is not a 7-th power, then the function $C_u$ yields the lowest possible linearity that a quadratic function can achieve, unless it is APN. Based on experimental observations, we conjecture that $C_u$ is neither a permutation (unless $u=0$) nor APN for any $m>3$. 

In the last part of the paper, we explore the CCZ-equivalence classes (named in short CCZ-classes in the following) of the known quadratic APN permutations in dimension 9, i.e., the Gold APN permutations $x \mapsto x^d,d \in \{3,5,17\}$ as well as $F_0$ and $F_1$. By utilizing EA-equivalence class invariants based on the algebraic degree and the thickness of a vector space, we observe that $F_0$ and $F_1$ consist of at least 12 and 19 distinct EA-equivalence classes (named in short EA-classes in the following), respectively. Interestingly, for $F_0$ (resp., $F_1$) there are at least 6 (resp., 8) distinct EA-classes that contain permutations. Similarly as it was shown for Gold APN permutations in dimension $n$ being a multiple of 3, we observe that a permutation EA-inequivalent to $F_0$ and $F_0^{-1}$ (resp., $F_1$ and $F_1^{-1}$) can be constructed from $F_0$ (resp., $F_1$) by just applying EA transformations and inversion. In particular, for $m=3$ and $u \in \F_{2^3} \setminus \{0,1\}$, we show that $C_u^{-1}+(x+x^4,0,0)$ is a permutation such that $\left(C_u^{-1}+(x+x^4,0,0)\right)^{-1}$ is EA-inequivalent to both $C_u$ and $C_u^{-1}$. For this property, the condition $m=3$ is necessary in the sense that, if $C_u$ is a permutation, then $C_u^{-1}+(x+x^4,0,0)$ is only a permutation if $m=3$.

\section{Preliminaries}
We now recall the important terminology and results needed in this work. For more details, we refer to~\cite{carlet_2021}. We study functions between finite-dimensional $\F_2$-vector spaces. First, we recall the notions of two well-known equivalence relations between such functions.
In the following, let $\mathcal{V}, \mathcal{W}, \mathcal{V'}, \mathcal{W'}$ be finite-dimensional $\F_2$-vector spaces. Two functions $F \colon \mathcal{V} \rightarrow \mathcal{W}$ and $F' \colon \mathcal{V'} \rightarrow \mathcal{W'}$ are called \emph{extended-affine equivalent} (or \emph{EA-equivalent} for short), if there exist affine bijections $A_1 \colon \mathcal{V'} \rightarrow \mathcal{V}, A_2 \colon \mathcal{W} \rightarrow \mathcal{W'}$ and an affine function $A_3 \colon \mathcal{V'} \rightarrow \mathcal{W'}$ such that $F' = A_2 \circ F \circ A_1 + A_3$. The functions $F$ and $F'$ are called \emph{CCZ-equivalent}~\cite{DBLP:journals/dcc/CarletCZ98}, if there exists an affine bijection $\sigma \colon \mathcal{V} \times \mathcal{W} \rightarrow \mathcal{V'} \times \mathcal{W'}$ such that $\sigma\left(\{(x,F(x)) \mid x \in \mathcal{V}\}\right) = \{(x,F'(x)) \mid x \in \mathcal{V'}\}$. The notion of CCZ-equivalence generalizes the notion of  EA-equivalence in the sense that two EA-equivalent functions are also CCZ-equivalent. Note that CCZ-equivalence is strictly more general than EA-equivalence combined with taking inverses in the case of permutations~\cite{DBLP:journals/tit/BudaghyanCP06}.

Let $d \in \mathbb{N}\coloneqq \{1,2,\dots,\}$. For a function $F \colon \mathcal{V} \rightarrow \mathcal{W}$ and a vector $\alpha \in \mathcal{V}$, the \emph{first-order derivative in direction $\alpha$} is defined as the function $\Delta_{\alpha}F\colon \mathcal{V} \mapsto \mathcal{W}, x \mapsto F(x)+F(x+\alpha)$. The \emph{differential uniformity}~\cite{DBLP:conf/eurocrypt/Nyberg93} of $F$ is defined as $D(F) \coloneqq \max_{\alpha \in \mathcal{V} \setminus \{0\},\beta \in \mathcal{W}}|\{x \in \mathcal{V} \mid \Delta_{\alpha}F(x)=\beta\}|$. If $D(F)=d$, we also say that $F$ is \emph{differentially $d$-uniform}. Differentially 2-uniform functions achieve the lowest possible differential uniformity and are called \emph{almost perfect nonlinear} (or \emph{APN} for short)~\cite{DBLP:conf/crypto/NybergK92}.

Let $\langle \cdot, \cdot \rangle_{\mathcal{V}} \colon \mathcal{V} \times \mathcal{V} \rightarrow \F_2$ and $\langle \cdot, \cdot \rangle_{\mathcal{W}} \colon \mathcal{W} \times \mathcal{W} \rightarrow \F_2$ be non-degenerate symmetric bilinear forms (i.e., inner products). For a function $F \colon \mathcal{V} \rightarrow \mathcal{W}$, any function $\mathcal{V} \rightarrow \F_2, x \mapsto \langle b,F(x) \rangle_{\mathcal{W}}$, where $b \in \mathcal{W}$, is called a \emph{component} of $F$. We call a component \emph{non-trivial} if $b \neq 0$.  The \emph{Walsh transform} of $F \colon \mathcal{V} \rightarrow \mathcal{W}$ at $(\alpha,\beta) \in \mathcal{V} \times \mathcal{W}$ is defined as 
\[ \widehat{F}(\alpha,\beta) = \sum_{x \in \mathcal{V}}(-1)^{\langle \alpha,x\rangle_{\mathcal{V}}+\langle \beta,F(x)\rangle_{\mathcal{W}}}\]
and the \emph{linearity} of $F$ is defined as $\Lin{(F)} \coloneqq \max_{\alpha \in \mathcal{V},\beta \in \mathcal{W}\setminus \{0\}} |\widehat{F}(\alpha,\beta)|$. It is well known that both the differential uniformity as well as the linearity of $F$ are invariant under CCZ-equivalence. Note that,  if $\dim(\mathcal{V}) = \dim(\mathcal{W}) = n$, the linearity of $F$ is lower bounded by $2^{\frac{n+1}{2}}$ and $F$ is called \emph{almost bent} if it matches this lower bound by equality~\cite{DBLP:conf/eurocrypt/ChabaudV94}.

Throughout this work, for $m \in \mathbb{N}$, let $\F_{2^m}$ denote the finite field with $2^m$ elements. Each $m$-dimensional $\F_2$-vector space can be equipped with a multiplication and be interpreted as $\F_{2^m}$. In this work, we concentrate on the vector spaces $\mathcal{V} =\mathcal{W} = (\F_{2^m})^t$ for $m,t \in \mathbb{N}$. In this case, a function $F \colon \mathcal{V} \rightarrow \mathcal{W}$ is said to be in \emph{$t$-variate representation} (for $t=1,2,3$, instead of $t$-variate, we say \emph{univariate}, \emph{bivariate}, and \emph{trivariate}, respectively).  Let \[\trace{}\colon \F_{2^m} \rightarrow \F_2, \quad x \mapsto \sum_{i=0}^{m-1}x^{2^i}\] denote the \emph{absolute trace function} over $\F_{2^m}$. For $\alpha_1,\dots,\alpha_t,x_1,\dots,x_t \in \F_{2^m}$, we use the bilinear form defined by $\langle(\alpha_1,\dots,\alpha_t),(x_1,\dots,x_t) \rangle_{\mathcal{V}} = \sum_{i=1}^t\trace{(\alpha_i x_i)}$ as the inner product over $\mathcal{V}$.

A function $F \colon (\F_{2^m})^t \rightarrow (\F_{2^m})^t$ can be given by $t$ \emph{coordinate functions} $f_1,\dots,f_t \colon (\F_{2^m})^t \rightarrow \F_{2^m}$ as $F(x_1,\dots,x_t) = (f_1(x_1,\dots,x_t),\dots,f_t(x_1,\dots,x_t))$. 
Each coordinate function $f_i$ can be uniquely expressed as a multivariate polynomial in $\F_{2^m}[X_1,\dots,X_t]/(X_1^{2^m}+X_1,\dots,X_t^{2^m}+X_t)$ via \[f_i \colon (\F_{2^m})^t \rightarrow \F_{2^m}, \quad x \mapsto \sum_{j_1=0}^{2^m-1}\dots \sum_{j_t=0}^{2^m-1}\left(\omega_{j_1,\dots,j_t}\prod_{i=1}^t x_i^{j_i}\right), \quad \omega_{j_1,\dots,j_t} \in \F_{2^m}.\]
The \emph{algebraic degree} of the coordinate function $f_i$ is then defined as \[\max_{\{(j_1,\dots,j_t) \mid \omega_{j_1,\dots,j_t} \neq 0\}} \left( \wt(j_1) + \dots + \wt(j_t)\right),\] where $\wt(k)$ denotes the Hamming weight of the binary expansion of $k \in \mathbb{N}$. The \emph{algebraic degree of $F$} is defined as the maximum algebraic degree of all its coordinate functions. Functions with algebraic degree equal to 2 are called \emph{quadratic} and functions with algebraic degree at most 1 are called \emph{affine}.

A vector subspace $\mathcal{U}$ of a vector space $\mathcal{V}$ has an \emph{orthogonal}, denoted $\mathcal{U}^{\perp}$. It is the set of all elements $x \in \mathcal{V}$ such that $\langle x,u\rangle_{\mathcal{V}} = 0$ for all $u \in \mathcal{U}$. 

\section{Decomposition}
\label{sec:decomposition}
In this section, we outline our approach that allowed us to obtain the trivariate representation of the two sporadic APN permutations in dimension 9.

\subsection{Generalizing the TU-decomposition}
Our approach for decomposing these functions is based on the \emph{TU-decomposition}, as first identified in~\cite{DBLP:conf/eurocrypt/BiryukovPU16}, which was later formalized as being a particular case of \emph{$t$-twisting}~\cite{DBLP:journals/ffa/CanteautP19}. The goal of this line of work is as follows. Let $F : \F_2^n \to \F_2^n$ be a function that we want to \emph{decompose}. In practice, this process consists in writing the space $\F_2^n$ as a Cartesian product $\F_2^t \times \F_2^{n-t}$, and then in identifying linear bijections $\eta : \F_2^t \times \F_2^{n-t} \to \F_2^n$ and $\mu : \F_2^n \to \F_2^t \times \F_2^{n-t}$ such that the function
\begin{equation*}
    G : 
    \begin{cases}
        \F_2^t \times \F_2^{n-t} &\to \F_2^t \times \F_2^{n-t} \\
        (x,y) &\mapsto \mu \circ F \circ \eta (x,y)
    \end{cases}
\end{equation*}
has a ``nice'' expression. In practice, we write $G$ as $G : (x,y) \mapsto (U_y(x), T_x(y))$, where $T_x$ maps $\F_2^t$ to itself for all $x \in \F_2^{n-t}$, and where $U_y$ maps $\F_2^{n-t}$ to itself for all $y \in \F_2^t$. These bivariate functions $T$ and $U$ give its name to the TU-decomposition. While we can always perform such a partition of the input and output space of a function, the term ``TU-decomposition'' refers to the case where $T_x$ is a permutation for all $x \in \F_2^{n-t}$. While it is one ``nice'' property, we usually want that $T$ and $U$ have very sparse algebraic representations when interpreting $\F_2$-vector spaces as finite fields of characteristic 2.

The property of $T$ in a TU-decomposition can be generalized as follows.
\begin{definition}[Permutation-concatenation]
    Let $\mathcal{V}$ be a finite-dimensional $\F_2$-vector space, $F : \mathcal{V} \to \mathcal{V}$ be a function, $\mathcal{U}$ be a subspace of $\mathcal{V}$, and $\rho_{\mathcal{U}} : \mathcal{V} \to \mathcal{U}$ be the linear projection which is the identity on $\mathcal{U}$. We say that $F$ is a \emph{permutation-concatenation of $\mathcal{U}$} if, for all $y \in \mathcal{V}$, the function 
    \begin{equation*}
        G_y :
        \begin{cases}
            \mathcal{U} &\to \mathcal{U} \\
            x &\mapsto \rho_{\mathcal{U}} \circ F (y + x)
        \end{cases}
    \end{equation*}
    is a permutation of $\mathcal{U}$.
\end{definition}

To investigate such patterns, we first need the notion of \emph{Walsh zeroes}.
\begin{definition}[Walsh zeroes~\cite{DBLP:journals/ffa/CanteautP19}]
    Let $\mathcal{V}$ and $\mathcal{W}$ be finite-dimensional $\F_2$-vector spaces. The \emph{Walsh zeroes} of a function $F \colon \mathcal{V} \rightarrow \mathcal{W}$, denoted by $\mathcal{Z}_F$, is the set of the coordinates of the zeroes in its Walsh spectrum together with $(0,0)$, i.e. 
    \begin{equation*}
        \mathcal{Z}_F = \left\{ (a,b) \in \mathcal{V} \times \mathcal{W} \mid \walshC{F}(a,b)=0 \right\} \cup (0,0) ~.
    \end{equation*}
\end{definition}

Let us generalize the notion of the TU-decomposition so as to allow an arbitrary number of ``sub-blocks'' rather than only two.
First, let us formalize this notion of ``blocks''. In the original TU-decomposition, we split the input space $\F_2^n$ into $\F_2^{n-t} \times \F_2^t$, and the ``$T$'' part is a permutation-concatenation of $\F_2^t$.

However, we can consider a more general case captured by the following definition. For a set $X$ we denote by $\mathcal{P}(X)$ the \emph{power set} of $X$, i.e., the set of all subsets of $X$.
\begin{definition}
  Let $\mathcal{V}$ be a finite-dimensional $\F_2$-vector space. We call \emph{block partition} of $\mathcal{V}$ a set $\{B_{i}\}_{1 \leq i \leq \ell} \subseteq \mathcal{P}(\mathcal{V})$ of subspaces of $\mathcal{V}$ such that $\cup_{i}B_{i}$ spans $\mathcal{V}$ and such that, for all $i \neq j$, the space $B_{i}$ is orthogonal to $B_{j}$ (i.e., for all $x \in B_i$ and $y \in B_j$, we have $\scalarprodV{x}{y} = 0$).
\end{definition}
We remark that the properties of a block partition $\{ B_{i} \}_{1 \leq i \leq \ell}$ of $\mathcal{V}$ impose that $\sum_{i}\dim(B_{i})=\dim(\mathcal{V})$, so the space $B_1 \times B_2 \times \dots \times B_{\ell}$ is isomorphic to the internal direct sum $B_1 \oplus B_2 \oplus \dots \oplus B_{\ell} = \mathcal{V}$.

As we can see, for any fixed $n,t \in \mathbb{N}$ with $t \leq n$, the set $ \mathcal{B} = \{B_1,B_2\}$ defined by $B_1 = \{(x_1,x_2,\dots,x_n) \in \F_2^n \mid x_{t+1} = x_{t+2} = \dots = x_n = 0 \}$ and $B_2 = \{(x_1,x_2,\dots,x_n) \in \F_2^n \mid x_{1} = x_{2} = \dots = x_{t} = 0 \}$ is a block partition of $\F_{2}^{n}$ (note that $B_1$ is isomorphic to $\F_t$ and $B_2$ is isomorphic to $\F_2^{n-t}$, so by a slight abuse of notation, we also write $\mathcal{B} = \{\F_2^t,\F_2^{n-t}\}$).  Unlike in the ``classical'' TU-decomposition case, there is no reason to restrict ourselves to the case $\ell=2$.

If $\mathcal{B} = \{ B_{i} \}_{1 \leq i \leq \ell}$ is a block partition of $\mathcal{V}$, we associate to it a set of linear projections $\{ \rho_{i} \}_{1 \leq i \leq \ell}$ such that, for all $1 \leq i \leq \ell$, we have $\rho_{i}(\mathcal{V}) = B_{i}$. We then have that
\begin{equation*}
  \mu : 
  \begin{cases}
    \mathcal{V} &\to B_1 \times \dots \times B_{\ell} \\
    x &\mapsto \big(\rho_{1}(x), ..., \rho_{\ell}(x)\big)
  \end{cases}
\end{equation*}
is an isomorphism (and $x = \sum_{i=1}^{\ell}\rho_i(x))$. In what follows, we will often implicitly rely on such a mapping $\mu$ by writing $x_{i}$ instead of $\rho_{i}(x)$. In the case of a TU-decomposition, we have that $T_y(x) = \rho_1 \circ F(x \| y)$, and that $U_x(y) = \rho_2 \circ F(x \| y)$.

\begin{theorem}
  \label{thm:tu-star}
  Let $\mathcal{V}$ be a finite-dimensional $\F_2$-vector space and let $F : \mathcal{V} \to \mathcal{V}$ be a function, $\mathcal{Z}_{F}$ be its Walsh zeroes, and let $\mathcal{B} = \{ B_{i} \}_{1 \leq i \leq \ell}$ be a block partition of $\mathcal{V}$. For every $i\in \{1,\dots ,\ell\}$, the function $\rho_{i} \circ F$ is a permutation-concatenation of $B_i$ if and only if $B_{i}^{\perp} \times B_{i} \subseteq \mathcal{Z}_{F}$.
\end{theorem}

Proposition~6 of~\cite{DBLP:journals/ffa/CanteautP19} is a particular case of
Theorem~\ref{thm:tu-star} which corresponds to $\mathcal{V} = \F_2^n$ and
$\mathcal{B} = \{\F_{2}^{t}, \F_{2}^{n-t}\}$.
Thus, it should come as no surprise that our proof has a similar structure to that in~\cite{DBLP:journals/ffa/CanteautP19}.
\begin{proof}
Let $i \in \{1,\dots ,\ell\}$. We first observe that $B_i \cup B_i^{\top}$ spans $ \mathcal{V}$. Indeed, by the definition of a block partition, we have that $B_i \cup \bigcup_{j \in \{1,\dots,\ell\}, j \neq i}B_j$ spans $\mathcal{V}$ and $\bigcup_{j \in \{1,\dots,\ell\}, j \neq i}B_j \subseteq B_i^{\top}$.
  Let us now fix an element $(a,b) \in B_i^{\perp} \times B_i$. First of all, for the Walsh transform of $F$, we have 
  \begin{equation*}
    \begin{split}
      \walshC{F}(a, b)
      &= \sum_{x \in \mathcal{V}} (-1)^{\scalarprodV{a}{x} + \scalarprodV{b}{F(x)}} \\
      &= \sum_{x \in \mathcal{V}} (-1)^{\scalarprodV{a}{x'}  +  \scalarprodV{b}{\rho_{i} \circ F(x'+x_i)}} ~,
    \end{split}      
  \end{equation*}
  where  $x = x'+x_{i}$ with $x_i \in B_i$ and $x' \in B_i^{\perp}$. We obtain
  \begin{equation*}
    \begin{split}
      \walshC{F}(a, b)
      &= \sum_{x' \in B_{i}^{\perp}, x_{i} \in B_{i}} (-1)^{\scalarprodV{a}{x'} + \scalarprodV{b}{\rho_{i} \circ F(x'+x_{i})}} \\
      &= \sum_{x' \in B_{i}^{\perp}} (-1)^{\scalarprodV{a}{x'}} \sum_{x_{i} \in B_{i}} (-1)^{\scalarprodV{b}{\rho_{i} \circ F(x'+x_{i})}}
    \end{split}      
  \end{equation*}
  This sum can be seen as the Fourier transform in $a$ of the integer valued function $g_b : B_i^{\perp} \to \mathbb{Z}, g_b : x' \mapsto \sum_{x_{i} \in B_{i}} (-1)^{\scalarprodV{b}{\rho_{i} \circ F(x'+x_{i})}}$. As a consequence, for a given $b \in B_i \setminus \{0\}$, $\walshC{F}(a,b)$ is equal to zero for all $a \in B_i^{\perp}$ if and only if $\walshC{g_b}(a) = 0$ for all $a \in B_i^{\perp}$. For $b=0$ and $a \neq 0$, it anyway holds that $\walshC{F}(a,b) = \walshC{g_b}{(a)} = 0$.
  As the only function $g$ for which $\walshC{g}(a)=0$ for all $a \in B_i^{\perp}$ is the all-zero function, we deduce that $B_{i}^{\perp} \times B_{i} \subseteq \mathcal{Z}_{F}$ if and only if, for all $b \in B_i \setminus \{0\}$, we have that $g_b$ is the all-zero function.  In turn, this function is constant and equal to zero for all $b \in B_i \setminus \{0\}$ if and only if all the non-trivial components of $x' \mapsto \rho_{i} \circ F (x' + x_{i})$ are balanced, meaning that $x' \mapsto \rho_{i} \circ F (x' + x_{i})$ is a bijection. We deduce that $B_{i}^{\perp} \times B_{i}$ is in $\mathcal{Z}_{F}$ if and only if $\rho_{i} \circ F$ is a permutation-concatenation of $B_{i}$.
\end{proof}

The main advantage of this alternative view of the TU-decomposition is that it can trivially handle the case where a function consists of the concatenation of multiple permutation-concatenations, while the original framework of the Proposition~6 of~\cite{DBLP:journals/ffa/CanteautP19} could only recover one permutation-concatenation. Indeed, the existence of two distinct vector spaces of zeroes is not sufficient to deduce that $F$ has two permutation-concatenations, but below we will show how to do it using Theorem~\ref{thm:tu-star}.

However, an advantage of the proposition in~\cite{DBLP:journals/ffa/CanteautP19} is that it can easily handle the case where the input and output sizes are different for $F$, which is not the case for Theorem~\ref{thm:tu-star}. It also does not require that the space $B$ over which there is a permutation-concatenation is such that $B^{\perp} \cup B$ spans the whole space; a property which is implied in the case of Theorem~\ref{thm:tu-star} by the properties of a block partition.

\subsection{Practical Usage}

Let $F$ be a function operating on a vector space $\mathcal{V}$ over $\F_2$, let $\mathcal{Z}_F$ be its Walsh zeroes, and let $\mathcal{B} = \{B_i\}_{1 \leq i \leq \ell}$ be a block partition of $\mathcal{V}$. We also denote $n = \dim{(\mathcal{V})}$. Using Theorem~\ref{thm:tu-star}, we can say that $F$ is a permutation-concatenation over blocks $\{B_j\}_{j \in J}$ for some $J \subseteq \{1,...,\ell\}$ if and only if $B_j^{\perp} \times B_j$ is in $\mathcal{Z}_F$ for all $j \in J$.

By definition, the elements of $\mathcal{Z}_F$ are pairs $(x,y) \in \mathcal{V} \times \mathcal{V}$. The two sides of such pairs do not play symmetric roles. For instance, the subspace $\{(x,0) \mid x \in \mathcal{V}\}$ is always in the Walsh zeroes of a function, while $\{(0,x) \mid x \in \mathcal{V}\}$ is in it if and only if the function is a permutation. Furthermore, the right-hand side coordinate of a subspace of $\mathcal{Z}_F$ has a particular interaction with EA-equivalence which we summarize below.

Let $F$ and $G$ be EA-equivalent functions of $\mathcal{V}$, so that $G(x) = (\mu \circ F \circ \eta)(x) + \phi(x)$, where $\eta$ and $\mu$ are linear\footnote{If they are affine, then their linear part should be used in the following instead. Similarly, if $\phi$ is affine then we use its linear part instead. In general, the set $\mathcal{Z}_G$ is invariant if we compose $G$ with simple translations, and thus our results are independent from said translations.} permutations, and where $\phi$ is a linear function. As a consequence, the graphs of these functions are related as follows:
\begin{equation*}
    \big\{(x, G(x)) \mid x \in \mathcal{V} \big\}
    ~=~
    \left[\begin{array}{cc}
    \eta^{-1} & 0 \\
    \phi \eta^{-1} & \mu
    \end{array}\right]
    \big\{(x, F(x)) \mid x \in \mathcal{V} \big\} ~,
\end{equation*}
as evidenced by a simple change of variable $y \gets \eta^{-1}(x)$ in the right hand side of the equation that rewrites the pair $\left(\eta^{-1}(x), \mu \circ F(x) + \phi \circ \eta^{-1} (x)\right)$ as $\left(y, \mu \circ F \circ \eta(y) + \phi(y)\right) = (y, G(y))$. Lemma~2 of~\cite{DBLP:journals/ffa/CanteautP19} states that if the graph of $G$ is the image of $F$ by a linear permutation $\mathcal{L}$, then $\mathcal{Z}_G = (\mathcal{L}^\top)^{-1}(\mathcal{Z}_F)$, which in our case yields
\begin{equation*}
    \mathcal{Z}_G
    ~=~
    \left[\begin{array}{cc}
    ({\eta^{-1}})^\top & (\phi {\eta^{-1}})^\top \\
    0 & \mu^\top
    \end{array}\right]^{-1}
    (\mathcal{Z}_F) ~.
\end{equation*}

This implies that if a vector space $\{(f_i,e_i)\}_{1 \leq i \leq 2^{n}}$ is contained in $\mathcal{Z}_F$, then a space of the form $\{(g_i,e'_i)\}_{1 \leq i \leq 2^{n}}$ with $e'_i = (\mu^\top)^{-1}(e_i)$ is contained in $\mathcal{Z}_G$. In particular, if several $n$-dimensional subspaces contained in $\mathcal{Z}_F$ are such that their right hand sides have a pairwise intersection reduced to $\{ 0 \}$, then $\mathcal{Z}_G$ will contain vector spaces with the same property. Furthermore, by considering the right-hand side of the spaces contained in $\mathcal{Z}_F$ and $\mathcal{Z}_G$, we can recover some information about $\mu$, namely that it maps $e_i$ to $e'_i$.

To better use this observation, we recall the notions of \emph{thickness} and \emph{thickness spectrum} as they will play a crucial role in determining suitable dimensions for the block partitions.
\begin{definition}[Thickness spectrum~\cite{DBLP:journals/ffa/CanteautP19}]
    Let $\mathcal{Z}_F$ be the Walsh zeroes of $F \colon \mathcal{V} \rightarrow \mathcal{W}$, $n = \dim(\mathcal{V})$,
    and let $\Sigma_F \subseteq \mathcal{P}(\mathcal{Z_F})$ be the set of vector spaces of dimension $n$ in $\mathcal{Z}_F$. The \emph{thickness spectrum} of $F$ is the multiset $\{t_i, n_{t_i}\}_{t_i\geq 0,n_{t_i}>0}$ such that, for all $i$, we have $|\{U \in \Sigma_F \mid t(V) = t_i \}| = n_i$, where $t(U)$ denotes the dimension of the projection of the $n$-dimensional vector space $U \subseteq  \mathcal{V} \times \mathcal{W}$ on $\{(0,y) \in \mathcal{V} \times \mathcal{W} \mid y \in \mathcal{W} \}$. 
\end{definition}

The thickness spectrum of $F_0$ is $\{n_0=1, n_1=511, n_2=2590, n_3=1144, n_9=512\}$, meaning that its Walsh zeroes contain a total of 1144 vector spaces where the right hand coordinates of all elements yields a subspace of dimension $3$. Among those, we identified triples of spaces $(U_1,U_2,U_3)$ whose right hand sides are disjoint and span the full space $\F_2^9$. This is a strong indication that $F_0$ is EA-equivalent to a trivariate function from $(\F_{2^3})^3$ to itself where each coordinate is a permutation-concatenation. Indeed, such a trivariate representation is equivalent (up to affine-equivalence) to the existence of a block partition $\{B_1, B_2, B_3\}$ such that $F$ is a permutation-concatenation on all $B_i, i \in \{1,2,3\}$. However, we could only deduce the mapping $\mu$ from this observation.

Still, using one these three vector spaces (say, $U_1 = \left\{ \left(A(x,y), \mu^{-1}(y)\right) \mid (x,y) \in \F_2^6 \times \F_2^3 \right\}$ where $A$ is an affine function of rank $6$) it is possible to deduce  a first permutation-concatenation, i.e. a non-linear function $T : (x,y,z) \to T_{y,z}(x)$ that corresponds to the expression of the first coordinate of $F_0$ up to EA-equivalence. We found the permutations $T_{y,z}$ to all be identical up to the addition of $(y,z)$-dependent functions in the input and output. To put it differently, we observed that $T_{y,z}(x) = P(x + f(y,z)) + g(y,z)$. Furthermore, $P$ itself is linear equivalent to the cube function. Overall, once composed with well chosen linear permutations, we found $T$ to be a permutation-concatenation of the form $(x,y,z) \mapsto (x + \ell(y,z))^3 + q(y,z)$, where $\ell$ is a linear function and where $q$ is a quadratic function. This prompted us to brute-force functions composed of three components with such a structure. We further aided our search using some observations on the symmetries (in particular linear self-equivalences) of $F_0$, resp., $F_1$. Eventually, we found a trivariate permutation-concatenation $\phi_u$ such that, up to EA-equivalence, $F_0$ is equal to the function $(x,y,z) \mapsto \big(\phi_u(x,y,z), \phi_u(y,z,x),\phi_u(z,x,y)\big)$. The details of these functions are provided below. An identical approach for $F_1$ yielded an (almost) identical result.

\subsection{Trivariate Representations}
The two APN permutations can be decomposed into the APN permutations given by
\begin{equation}\label{eq:trivariate_decomposition} C_u \colon (\F_{2^m})^3 \rightarrow (\F_{2^m})^3, \quad (x,y,z) \mapsto (x^3+uy^2z, y^3+uxz^2,z^3+ux^2y)\;,\end{equation}
where $m=3$ and $u \in \F_{2^3}\setminus \{0,1\}$. If $u$ is a zero of $X^3+X+1$, then the permutation $C_u$ is EA-equivalent to $F_0$. Otherwise (i.e., if $u$ is a zero of $X^3+X^2+1$), it is EA-equivalent to $F_1$. In general, for each $m \in \mathbb{N}$ and $u \in \F_{2^m}$, the function $C_u$ has a lot of symmetries (i.e., self-equivalences):
\begin{itemize}
    \item The function $C_u$ is rotation symmetric, i.e., $C_u \circ r = r \circ C_u$, where $r : (x,y,z) \mapsto (y,z,x)$. Indeed, by defining the function $\phi_u \colon (\F_{2^m})^3 \to \F_{2^m}, (x,y,z) \mapsto x^3 + u y^2 z$, we can write $C_u$ as $(x,y,z) \mapsto (\phi_u(x,y,z), \phi_u(y,z,x),\phi_u(z,x,y))$.
    \item If we multiply each of the three input coordinates of $C_u$ by $\lambda \in \F_{2^m}$, each coordinate of the output is multiplied by $\lambda^3$. This property is reminiscent of the \emph{subspace property} of the Kim mapping, which was first identified by its inventors~\cite{Dillon-perm}, and later explained by its open butterfly structure~\cite{DBLP:conf/crypto/PerrinUB16}.
    \item As always for quadratic functions, adding a constant in the input of $C_u$ is the same as adding an affine function to the output of $C_u$.
\end{itemize}

\paragraph{Inverses.}
For $m=3$, the inverse of $C_u$ with $u \in \F_{2^3}$ being a zero of $X^3+X+1$ can be given as $C_u^{-1}\colon (x,y,z) \mapsto (\psi_u(x,y,z), \psi_u(y,z,x),\psi_u(z,x,y))$, where $\psi_u \colon (\F_{2^3})^3 \rightarrow \F_{2^3}$ is defined by 
\begin{align*} \psi_u(x,y,z)&=(u^2+u)yz^4 + (u^2+u)xy^5z^6 + (u^2+u+1)x^2y^2z + ux^4y^3z^5 \\ &+ x^5 + ux^5z^7 + (u^2+u+1)x^5y^7 + (u^2+1)x^6y^4z^2 + x^7yz^4.\end{align*}

Similarly, for $m=3$, the inverse of $C_u$ with $u \in \F_{2^3}$ being a zero of $X^3+X^2+1$ can be given as $C_u^{-1}\colon (x,y,z) \mapsto (\psi_u(x,y,z), \psi_u(y,z,x),\psi_u(z,x,y))$, where $\psi_u \colon (\F_{2^3})^3 \rightarrow \F_{2^3}$ is defined by 
\begin{align*} \psi_u(x,y,z)&=(u+1)yz^4 + (u^2+u+1)xy^5z^6 + x^2y^2z + ux^4y^3z^5 \\ &+ x^5 + ux^5z^7 + (u+1)x^5y^7 + (u^2+u)x^6y^4z^2 + (u^2)x^7yz^4.\end{align*}

It is straightforward to observe that both of those inverses are of algebraic degree 5.

\begin{remark}
Using the computer algebra system Magma~\cite{MR1484478}, the reader could verify the CCZ-equivalence of the trivariate functions $C_u$ for $m=3$ and $u \in \F_{2^3}\setminus \{0,1\}$ to $F_0$, resp., $F_1$ by utilizing the well-known method based on checking equivalence of linear codes, see~\cite[p.\@ 412]{carlet_2021}. The EA-equivalence between $C_u$ and $F_0$, resp., $F_1$ then follows because two quadratic APN functions are CCZ-equivalent if and only if they are EA-equivalent~\cite{yoshiara2012equivalences}.
\end{remark}

\section{Bounds on the Differential Uniformity and Linearity of $C_u$}
The property that $C_u$ with parameters $m=3$ and $u \in \F_{2^3}\setminus \{0,1\}$ is APN follows from the EA-equivalence to $F_0$, resp., $F_1$.
In this section, we study the differential uniformity and the linearity of the family of functions $C_u \colon (\F_{2^m})^3 \rightarrow (\F_{2^m})^3$ for $m$ being a multiple of three. As we will see, the case of $u$ not being a 7th-power is particularly interesting as it yields differentially $d$-uniform functions with $d\leq 8$ having low linearity. 

\subsection{On the Differential Uniformity of $C_u$}
In what follows, we will use the fact that $7 = 2^3-1$ divides $2^m-1$ if and only if $m$ is divisible by 3. As a consequence, we have $\gcd(m,7) \neq 1$ if and only if $m$ is divisible by 3 and in this case, $x \mapsto x^7$ is not a permutation, meaning that there are values $y$ in $\F_{2^m}$ such that, for all $x \in \F_{2^m}$, $x^7 \neq y$. We now prove the following result. 

\begin{theorem}
    Let $m$ be a multiple of 3 and let $\phi_u : (\F_{2^m})^3 \to \F_{2^m}$ be defined for some element $u \in \F_{2^m}$ as $\phi_u(x,y,z)=x^3 + u y^2 z$. If $u$ is not in the image of $x\mapsto x^7$, then \[C_u : (x,y,z) \mapsto (\phi_u(x,y,z), \phi_u(y,z,x), \phi_u(z,x,y))\] is a differentially $d$-uniform function with $d\leq 8$.
\end{theorem}
\begin{proof}
Let us fix an element $u \in \F_{2^m}$ which is not a 7-th power. To prove that $C_u$ is differentially $d$-uniform with $d\leq 8$, we will rely on the resolution of a system of differential equations. First, note that the first-order derivative of $\phi_u$ in direction $(\alpha,\beta,\gamma)$ can be given as
\begin{equation*}
    \Delta_{(\alpha, \beta, \gamma)}\phi_u(x,y,z) 
    ~=~
    \alpha x^2 + \alpha^2 x + u\gamma y^2 + u \beta^2 z + \alpha^3 + u \beta^2 \gamma ~.
\end{equation*}

In what follows, we cannot have that $\alpha = \beta = \gamma = 0$. In order to show that the differential uniformity of $C_u$ is bounded above by 8, we  prove that the homogeneous system associated to the equation $\Delta_{(\alpha,\beta,\gamma)}C_u(x,y,z) = (a,b,c)$, namely
\begin{equation}
    \label{sys:main-diff}
    \begin{cases}
        \alpha x^2 + \alpha^2 x + u\gamma y^2 + u\beta^2 z &= 0 \\
        \beta y^2 + \beta^2 y + u\alpha z^2 + u\gamma^2 x &=0 \\
        \gamma z^2 + \gamma^2 z + u\beta x^2 + u\alpha^2 y &= 0 \\
    \end{cases}
\end{equation}
has at most 8 solutions $(x,y,z) \in (\F_{2^m})^3$. We distinguish the following three cases:

\paragraph{Case with 2 zero input differences.}
Due to the rotation symmetry of $C_u$, we suppose without loss of generality that $\alpha \neq 0$ and that $\beta = \gamma = 0$. In this case, System~(\ref{sys:main-diff}) simplifies to
\begin{equation*}
    \begin{cases}
        \alpha x^2 + \alpha^2 x &= 0 \\
        u\alpha z^2 &= 0 \\
        u\alpha^2 y &= 0 \\
    \end{cases}
\end{equation*}
which cannot have more than 2 solutions $(x,y,z) \in (\F_{2^m})^3$, since $u\neq0$. 

\paragraph{Case with 1 zero input difference.}
Due to the rotation symmetry of $C_u$, we suppose without loss of generality that $\alpha \neq 0$, $\beta \neq 0$ and that $\gamma = 0$. In this case, System~(\ref{sys:main-diff}) simplifies to
\begin{equation*}
    \begin{cases}
        \alpha x^2 + \alpha^2 x + u\beta^2 z &= 0 \\
        \beta y^2 + \beta^2 y + u\alpha z^2  &= 0 \\
        \beta x^2 + \alpha^2 y &= 0 \\
    \end{cases}\;.
\end{equation*}
We then substitute $x$ with $\alpha x$ and $y$ with $\beta y$ to obtain
\begin{equation}
    \label{sys:2-zeroes}
    \begin{cases}
         x^2 + x  &= u\alpha^{-3}\beta^2 z \\
         y^2 + y   &= u\alpha\beta^{-3} z^2 \\
         x^2  &= y \\
    \end{cases}\;.
\end{equation}
This implies that $y$ can be replaced with $x^2$ in the second equation to obtain
\begin{equation*}
    x^2+ x ~=~ (u\alpha\beta^{-3} z^2)^{1/2} = u^{1/2}\alpha^{1/2}\beta^{-3/2} z ~.
\end{equation*}
Combining this equation with the first one in the previous system yields
\begin{equation*}
    u^{1/2}\alpha^{1/2}\beta^{-3/2} z = u \alpha^{-3}\beta^2 z ~.
\end{equation*}
If $z = 0$ then $(x,y) = (0,0)$ or $(x,y)=(1,1)$. Thus, the system has at most two solutions of the form $(x,y,0)$. Suppose now that $z \neq 0$. We then need that
\begin{equation*}
    u\alpha\beta^{-3} = u^2\alpha^{-6}\beta^4 ~,
\end{equation*}
which is equivalent to $u = (\alpha/\beta)^7$. This is a contradiction to $u$ not being a 7-th power. Thus, the system has at most two solutions $(x,y,z) \in (\F_{2^m})^3$.

\paragraph{Case with no zero input difference.}
Since $\alpha, \beta, \gamma$ are all non-zero, we could consider solutions of the form $\alpha x, \beta y, \gamma z$.
The homogeneous system associated to the equation $\Delta_{(\alpha,\beta,\gamma)}C_u(\alpha x,\beta y,\gamma z)=(a,b,c)$ writes: 
\begin{equation}\label{sys1}\left\{\begin{array}{l}x^2+x+f(y^2+z)=0\\y^2+y+g(z^2+x)=0\\z^2+z+h(x^2+y)=0\end{array}\right.\end{equation}
where \begin{equation}\label{fgh} f=u\alpha^{-3}\beta^2\gamma\neq 0,\; g=u\alpha \beta^{-3}\gamma^2\neq 0,\; h=u\alpha^2\beta\gamma^{-3}\neq 0\end{equation} and therefore $fgh=u^3$. Note that $f$, $g$ and $h$ are not any elements of $\mathbb{F}_{2^n}$ such that $fgh=u^3$, since for instance $gf^{-2}u$ is a 7-th power. The sum of $\frac 1f$ times the first equation, $\frac1g$ times the second equation and $\frac1h$ times the third equation provides the following condition: \[(1+\frac 1f)(x^2+x)+(1+\frac 1g)(y^2+y)+(1+\frac 1h)(z^2+z)=0\;.\]
At least one of the coefficients $(1+\frac 1f),(1+\frac 1g)$ and $(1+\frac 1h)$ is nonzero since if $\frac 1f=\frac 1g=\frac 1h=1$ then $\frac 1{fgh}=1$, that is $u^3=1$, a contradiction. Without loss of generality, assume that $1+\frac 1h\neq 0$, that is, $h\neq 1$. Then we can replace $z+z^2$ by its value from the last equation of the system and we obtain $(1+\frac 1f)(x^2+x)+(1+\frac 1g)(y^2+y)+(h+1)(x^2+y)=0$, that is: \[(h+\frac 1f)x^2+(1+\frac 1f)x=(1+\frac 1g)y^2+(h+\frac 1g)y\;.\]
Eliminating $z$ between the two first equations yields:
\[g[x^2+x+f(y^2+z)]^2+f^2[y^2+y+g(z^2+x)]=gx^4+gx^2+gf^2x+gf^2y^4+f^2y^2+f^2y=0\;.\]

We arrive then to the following system necessarily satisfied by $x$ and $y$:
\begin{equation}\label{sys2}\left\{\begin{array}{l}(h+\frac 1f)x^2+(1+\frac 1f)x=(1+\frac 1g)y^2+(h+\frac 1g)y
\\
gx^4+gx^2+gf^2x=gf^2y^4+f^2y^2+f^2y\end{array}\right.\end{equation}
and for every solution $(x,y) \in (\F_{2^m})^2$ of this system, there is a unique $z \in \F_{2^m}$ such that $(x,y,z)$ satisfies  the original system; this value of $z$ is determined by any one of its two first equations.

 Note that each of the two equations in System~(\ref{sys2}) is invariant when replacing $(x,y)$ by $(x+1,y+1)$; this comes from the fact that System~(\ref{sys1}) is invariant when replacing $(x,y,z)$ by $(x+1,y+1,z+1)$. This property allows to take as new variables $X=x^2+x$ and $S=x+y$ (which are both invariant when replacing $(x,y)$ by $(x+1,y+1)$). Then System~(\ref{sys2}) is equivalent to:
\begin{equation}\label{sys3}\left\{\begin{array}{l}(1+\frac 1g)S^2+(h+\frac 1g)S=(h+\frac 1f+1+\frac 1g)X
\\
gf^2S^4+f^2S^2+f^2S=g(1+f^2)X^2+f^2(g+1)X\\\trace(X)=0\end{array}\right.\;.\end{equation}
The function $C_u$ is differentially $d$-uniform with $d\leq 8$ if and only if, for every $f,g,h$ satisfying System~(\ref{fgh}), System~(\ref{sys3}) has at most four solutions $(S,X) \in (\F_{2^m})^2$. 

If $h+\frac 1f+1+\frac 1g = 0$, System (\ref{sys3}) has at most four solutions $(S,X)$ since the coefficients of $S^2$ and $S$ in the first equation, i.e., $1+\frac{1}{g}$ and $h+\frac{1}{g}$, cannot simultaneously be zero as otherwise $h$ would be 1. Therefore, let us assume $h+\frac 1f+1+\frac 1g \neq 0$. By eliminating $X^2$ and $X$ in the second equation of System (\ref{sys3}) by using the first equation yields a polynomial equation in $S$ of degree at most four, where the coefficient of $S^4$ is equal to $f^2gh^2 + \frac{1}{g} = (u^6+1)g^{-1}$. In particular, the coefficient of $S^4$ is non-zero since $u^6 \neq 1$.
\end{proof}

\begin{remark}
For $m\geq 3$, the property that $u \in \F_{2^m}$ is not a 7-th power is necessary for $C_u$ being differentially $d$-uniform with $d\leq 8$. Clearly, $u$ being nonzero is necessary. Suppose that $u \in \F_{2^m}\setminus \{0\}$ is a 7th-power and consider $\alpha = u^{\frac{1}{7}}, \beta = 1, \gamma = 0$.  Then, System~(\ref{sys:main-diff}) becomes
\begin{equation*}
    \begin{cases}
        \alpha x^2 + \alpha^2 x + \alpha^7 z &= 0 \\
        y^2 + y + \alpha^8 z^2  &=0 \\
        \alpha^7 x^2 + \alpha^9 y &= 0\;, \\
    \end{cases}
\end{equation*}
which is equivalent to $[y = \alpha^{-2}x^2] \wedge [z = \alpha^{-6}(x^2 + \alpha x)]$ and this system has $2^m$ solutions $(x,y,z)$. Note that for $m=3$, the only elements in $\F_{2^m}$ being a 7th-power are 0 and 1. The function $C_1 \colon \F_{2^3} \rightarrow \F_{2^3}$ is differentially $32$-uniform.
\end{remark}

\begin{remark}
If $m$ is even, $C_u$ cannot be a permutation. For example, we have $C_u(x,0,0) = (x^3,0,0)$ and the function $\F_{2^m} \rightarrow \F_{2^m}, x \mapsto x^3$ is 3-to-1 for even values of $m$. Further, if $m$ is arbitrary and $u \in \F_{2^m} \setminus \{0\}$ is a 7-th power, $C_u$ is also not a permutation. This can be seen by taking $\alpha \in \F_{2^m} \setminus \{0\}$ with $\alpha^7=u$ and considering the equation $\Delta_{(\alpha,1,0)}C_u(\alpha x,y,z) = (0,0,0)$, which is equivalent to the system 
\begin{equation*}
    \begin{cases}
        x + x^2 + \alpha^4 z &= 1 \\
        y + y^2 + \alpha^8 z^2  &=1 \\
        x^2 + y &= 1\;. \\
    \end{cases}
\end{equation*}
This system has a solution $(x,y,z)=(1,0,\alpha^{-4})$.
The case of $m$ being odd and $u \in \F_{2^m}$ not being a 7-th power is open. With the computer algebra system sage~\cite{sage}, we checked that, for all $m \in \{9,15,21\}$ and $u \in \F_{2^m}$ not being a 7-th power, the function $C_u$ is not a permutation. The check was performed by finding elements $(\alpha,\beta,\gamma) \neq (0,0,0)$ such that the equation $\Delta_{(\alpha,\beta,\gamma)}C_u(x,y,z) = (0,0,0)$ has a solution $(x,y,z)$.
\end{remark}

\begin{remark}
\label{rem:experiments}
It is open problem whether $C_u$ can be APN for $m>3$. In particular, we only found APN functions (more precisely, APN permutations) within the family $C_u$ for $m=3$. Note that for $m=6$, there exist differentially 4-uniform functions within the family $C_u$ (but no APN functions), see Table~\ref{tab:cu_6}. For all $m \in \{9,12,15,18\}$ and $u \in \F_{2^m}$ not being a 7-th power, we checked with the computer algebra system sage that $C_u$ is always differentially $8$-uniform. This check was performed by finding elements $(\alpha,\beta,\gamma) \neq (0,0,0)$ such that System~(\ref{sys:main-diff}) has 8 solutions $(x,y,z)$.
\end{remark}

It seems that the case of $m=3$ is very special for obtaining APN permutations within the family $C_u$ and we think that the small size of the field $\F_{2^3}$ is the key property, similarly as it was for the Butterfly family to contain APN permutations~\cite{DBLP:journals/ccds/CanteautPT19}. Therefore, we raise the following conjecture. 
\begin{conjecture}
\label{con:1}
Let $m> 3$ and let $u \in \F_{2^m}$. For $\phi_u : (\F_{2^m})^3 \to \F_{2^m}, \phi_u(x,y,z)=x^3 + u y^2 z$, let  \[C_u : (x,y,z) \mapsto (\phi_u(x,y,z), \phi_u(y,z,x), \phi_u(z,x,y))\;.\] Then, $C_u$ is not APN. Further, if $u \neq 0$,  $C_u$ is not a permutation.
\end{conjecture}

\begin{remark}
After the preprint of our work has been made public, Bartoli and Timpanella proved in~\cite{bartoli2022conjecture} that $C_u$ is not APN for $m>20$. Together with the experimental results described in Remark~\ref{rem:experiments}, this proves the first part of Conjecture~\ref{con:1}.
\end{remark}

\begin{table}
    \centering
    \caption{\normalsize\label{tab:cu_6}Properties of all $C_u \colon \F_{2^m} \rightarrow \F_{2^m}$ for $m=6$ and $u \in \F_{2^m}$ not being a 7-th power. $|\im(C_u)|$ denotes the size of the image of $C_u$, i.e., $|\{C_u(x) \mid x \in (\F_{2^m})^3\}|$.}
    \footnotesize
    \begin{tabular}{rccc}
        \toprule
        & minimal polynomial of $u$ & $D(C_u)$  & $|\im(C_u)|$ \\ \midrule
        1 & $X^6 + X^5 + X^4 + X^2 + 1$ & 4  & 77680 \\
        2 & $X^6 + X^4 + X^2 + X + 1$ & 4  & 76210 \\
        3 & $X^3 + X^2 + 1$ & 4 &  77680 \\
        4 & $X^3 + X + 1$ & 4 &  76210 \\ \midrule
        5 & $X^6 + X^4 + X^3 + X + 1$ & 8  & 74152 \\
        6 & $X^6 + X^5 + X^3 + X^2 + 1$ & 8  & 73564 \\
        7 & $X^6 + X^5 + X^2 + X + 1$ & 8  & 74152 \\
        8 & $X^6 + X^5 + X^4 + X + 1$ & 8  & 73564 \\
        9 & $X^6 + X^5 + 1$ & 8 &  74152 \\
        10 & $X^6 + X + 1$ & 8 &  73564 \\
        \bottomrule
    \end{tabular}
\end{table}

\subsection{On the Linearity of $C_u$}
To prove an upper bound on the linearity of $C_u$, we apply the approach described in~\cite{DBLP:journals/ffa/AnbarKM19}. In particular, we use the following result.
\begin{lemma}[Prop.\@ 2.4. of~\cite{DBLP:journals/ffa/AnbarKM19}]
\label{lem:common_zeros}
Let $k$ be an integer with $\gcd(k,m)=1$ and let $f_1, f_2$ be linearized polynomials of the form
\[a_0X + b_0Y + a_1X^{2^k} + b_1 Y^{2^k} + \dots + a_dX^{2^{dk}} + b_dY^{2^{dk}} \in \F_{2^m}[X,Y]\]
of degree $2^{d_1k}$ and $2^{d_2k}$, respectively. If $f_1$ and $f_2$ do not have a common factor, then $f_1$ and $f_2$ have at most $2^{d_1+d_2}$ common zeros.
\end{lemma}

By using a similar method as in~\cite{DBLP:journals/ffa/AnbarKM19}, we then obtain the following result for the linearity of $C_u$.

\begin{theorem}
\label{thm:linearity}
Let $m$ be a multiple of 3 and let $\phi_u : (\F_{2^m})^3 \to \F_{2^m}$ be defined for some element $u \in \F_{2^m}$ as $\phi_u(x,y,z)=x^3 + u y^2 z$. If $u$ is not in the image of $x\mapsto x^7$, then the linearity of \[C_u : (x,y,z) \mapsto (\phi_u(x,y,z), \phi_u(y,z,x), \phi_u(z,x,y))\] is bounded above as $\Lin{(C_u)} \leq  8^{1+\lfloor \frac{m}{2} \rfloor}$.
\end{theorem}
\begin{proof}
Since $C_u$ is quadratic, every non-trivial component function \[C_u^{(\alpha,\beta,\gamma)} \coloneqq \trace{(\alpha \phi_u(x,y,z) + \beta \phi_u(y,z,x) + \gamma \phi_u(z,x,y))}, \quad (\alpha,\beta,\gamma) \neq (0,0,0)\] of $C_u$ is plateaued, i.e., for all $S,T,U \in \F_{2^m}$ we have $\widehat{C_u}^{(\alpha,\beta,\gamma)}(S,T,U) \in \{0, \pm 2^{\frac{3m+d}{2}}\}$, where $d$ is the dimension of the linear space of $C_u^{(\alpha,\beta,\gamma)}$ (see~\cite[Prop.\@ 55]{carlet_2021}). We recall that the linear space of a Boolean function $f \in (\F_{2^m})^t \rightarrow \F_2$ is defined as
\[\LS(f) \coloneqq \{ \alpha \in (\F_{2^m})^t \mid \Delta_\alpha f \text{ is constant}\}\;,\]
which can be simplified for quadratic $f$ to $\LS(f)  = \{ \alpha \in (\F_{2^m})^t \mid \tilde{\Delta}_\alpha f = 0\}$,
where $\tilde{\Delta}_\alpha$ denotes the linear part of $\Delta_\alpha f$. Let us fix a non-trivial component $C_u^{(\alpha,\beta,\gamma)}$ of $C_u$. We are interested in the dimension $d$ of 
\[\LS(C_u^{(\alpha,\beta,\gamma)}) = \{ (S,T,U) \in (\F_{2^m})^3 \mid \tilde{\Delta}_{(S,T,U)}C_u^{(\alpha,\beta,\gamma)} = 0\}\;.\] We have
\begin{align*}
\tilde{\Delta}_{(S,T,U)}C_u^{(\alpha,\beta,\gamma)}(x,y,z) &= \trace{(\alpha S x^2 + \alpha S^2 x + \beta uU^2x + \gamma u T x^2)} \\
&+ \trace{(\beta T y^2 + \beta T^2 y + \gamma uS^2y + \alpha u U y^2)} \\
&+ \trace{(\gamma U z^2 + \gamma U^2 z + \alpha uT^2z + \beta u S z^2)} \\
&= \trace{(Ax^2)} + \trace{(By^2)} + \trace{(Cz^2)}
\end{align*}
with $A = \alpha S + \alpha^2 S^4 + \beta^2 u^2U^4 + \gamma u T$, $B = \beta T  + \beta^2 T^4 + \gamma^2 u^2S^4 + \alpha u U$, and $C = \gamma U + \gamma^2 U^4 + \alpha^2 u^2T^4 + \beta u S$. Clearly,  $\tilde{\Delta}_{(S,T,U)}C_u^{(\alpha,\beta,\gamma)}$ is the zero function if and only if $A = B = C = 0$. Thus, to obtain $|\LS(C_u^{(\alpha,\beta,\gamma)})|$, we need to determine the number of solutions $(S,T,U) \in (\F_{2^m})^3$ of the following system of linear equations:
\begin{equation}
\label{eq:linear_structures}
    \begin{cases}
         \alpha S + \alpha^2 S^4 + \beta^2 u^2U^4 + \gamma u T &= 0 \\
        \beta T  + \beta^2 T^4 + \gamma^2 u^2S^4 + \alpha u U &= 0 \\
        \gamma U + \gamma^2 U^4 + \alpha^2 u^2T^4 + \beta u S &= 0 ~.
    \end{cases}
\end{equation}
We again distinguish three cases depending on the number of nonzero values for $\alpha, \beta, \gamma$.

\paragraph{Case of 2 zero values.} Due to the rotation symmetry of $C_u$, we assume that $\alpha \neq 0$ and $\beta = \gamma = 0$. Then, System~(\ref{eq:linear_structures}) simplifies to 
\begin{equation*}
    \begin{cases}
         S(1 + \alpha S^3)  &= 0 \\
         \alpha u U &= 0 \\
         \alpha^2 u^2T^4  &= 0 ~,
    \end{cases}
\end{equation*}
which has either 4, 2, or 1 solution(s) $(S,T,U) \in (\F_{2^m})^3$, since $u\neq 0$.
\paragraph{Case of 1 zero value.} Due to the rotation symmetry of $C_u$, we assume that $\alpha \neq 0, \beta \neq 0$ and $\gamma = 0$. Then, System~(\ref{eq:linear_structures}) simplifies to 
\begin{equation*}
    \begin{cases}
         \alpha S + \alpha^2 S^4 + \beta^2 u^2U^4  &= 0 \\
        \beta T  + \beta^2 T^4  + \alpha u U &= 0 \\
       \alpha^2 u^2T^4 + \beta u S &= 0 ~.
    \end{cases}
\end{equation*}
Eliminating $S$ in the first equation yields
\begin{equation*}
    \begin{cases}
         \alpha^3 \beta^3 u^5 T^4 + \alpha^{10}u^8 T^{16} + \beta^6 u^6U^4  &= 0 \\
        \beta T  + \beta^2 T^4  + \alpha u U &= 0 \\
       \alpha^2 u^2T^4 + \beta u S &= 0 ~
    \end{cases}
\end{equation*}
and further eliminating $U$ in the first equation yields
\begin{equation*}
    \begin{cases}
         (\alpha^{14}u^8 + \beta^{14}u^2)T^{16} + (\alpha^7 \beta^3u^5 + \beta^{10} u^2)T^4   &= 0 \\
        \beta T  + \beta^2 T^4  + \alpha u U &= 0 \\
       \alpha^2 u^2T^4 + \beta u S &= 0 ~.
    \end{cases}
\end{equation*}
This system has at most 4 solutions as long as the coefficients of $T^{16}$ and $T^4$ in the first equation are not both zero. One can see that those coefficients are both zero if and only if $u=0$ or $u^3 = \left(\frac{\beta}{\alpha}\right)^7$, so $u^3$ must be a 7-th power. This is a contradiction to $u$ not being a 7-th power.

\paragraph{Case of no zero value.} We now consider the case of $\alpha \neq 0, \beta \neq 0$, and $\gamma \neq 0$. Eliminating $S^4$ in the second equation by means of the first and then eliminating $S$ by means of the third equation yields
\begin{equation*}
    \begin{cases}
         \alpha S + \alpha^2 S^4 + \beta^2 u^2U^4 + \gamma u T &= 0 \\
        (\alpha^2 \beta^3 + \alpha^3\gamma^2 u^3) T^4  + (\alpha^2 \beta^2 + \beta \gamma^3 u^3) T + (\alpha \gamma^4 u + \beta^3 \gamma^2 u^4) U^4 + (\alpha^3 \beta u + \alpha \gamma^3 u) U &= 0 \\
        \gamma U + \gamma^2 U^4 + \alpha^2 u^2T^4 + \beta u S &= 0 ~.
    \end{cases}
\end{equation*}

Now, eliminating $S$ in the first equation by using the third equation yields
\begin{equation}
\label{eq:linear_structures_2equations}
    \begin{cases}
        \alpha^{10}u^8T^{16} + \alpha^3 \beta^3 u^5 T^4 + \beta^4 \gamma u^5 T +  \\
        \alpha^2 \gamma^8 U^{16}  + (\alpha^2 \gamma^4 + \beta^6 u^6 + \alpha \beta^3 \gamma^2 u^3) U^4 + \alpha \beta^3 \gamma u^3 U &= 0 \vspace{.5em}\\
        (\alpha^2 \beta^3 + \alpha^3\gamma^2 u^3) T^4  + (\alpha^2 \beta^2 + \beta \gamma^3 u^3) T + (\alpha \gamma^4 u + \beta^3 \gamma^2 u^4) U^4 + (\alpha^3 \beta u + \alpha \gamma^3 u) U &= 0 ~
    \end{cases}
\end{equation}
for the first two equations of System~(\ref{eq:linear_structures}). The value of $S$ can be uniquely recovered from the solutions $(U,T) \in (\F_{2^m})^2$ of System~(\ref{eq:linear_structures_2equations}).

If $m$ is odd, then the two polynomials in $\F_{2^m}[T,U]$ on the left-hand side of System~(\ref{eq:linear_structures_2equations}) are of the form as in Lemma~\ref{lem:common_zeros} with $k=2$ and $d_1 = 2, d_2 = 1$. If $m$ is even, then the two polynomials on the left-hand side of System~(\ref{eq:linear_structures_2equations}) are of the form as in Lemma~\ref{lem:common_zeros} with $k=1$ and $d_1 = 4, d_2 = 2$. Thus, we need to show that the two polynomials do not have a common factor to deduce that $\Lin(C_u) \leq 2^{\frac{3m+3}{2}} = 8^{1+\lfloor \frac{m}{2} \rfloor}$ if $m$ is odd and $\Lin(C_u) \leq 2^{\frac{3m+6}{2}}=8^{1+\lfloor \frac{m}{2} \rfloor}$ if $m$ is even.

For a fixed $T \in \F_{2^m}$, let us consider the following two polynomials $p_T,q_T \in \F_{2^m}[U]$.
\begin{align*}
    p_T &= aU^{16} + bU^4 + cU + d_T \\
    q_T &= eU^4 + fU + g_T\;,
\end{align*}
where $a = \alpha^2 \gamma^8$, $b = \alpha^2 \gamma^4 + \beta^6 u^6 + \alpha \beta^3 \gamma^2 u^3$, $c = \alpha \beta^3 \gamma u^3$, $d_T = \alpha^{10}u^8T^{16} + \alpha^3 \beta^3 u^5 T^4 + \beta^4 \gamma u^5 T$, and $e = \alpha \gamma^4 u + \beta^3 \gamma^2 u^4$, $f = \alpha^3 \beta u + \alpha \gamma^3 u$, $g_T = (\alpha^2 \beta^3 + \alpha^3\gamma^2 u^3) T^4  + (\alpha^2 \beta^2 + \beta \gamma^3 u^3) T$. To show that the two polynomials in $\F_{2^m}[T,U]$ on the left hand side of System~(\ref{eq:linear_structures_2equations}) do not have a common non-constant factor, we need to show that the resultant (as a polynomial in $T$) of the two polynomials $p_T$ and $q_T$ is not constant zero (see, e.g.,~\cite[Theorem 2.20]{hirschfeld}). Note that the resultant $\resultant(p_T,q_T)$ of $p_T$ and $q_T$ is defined as the determinant of the Sylvester matrix $\sylvester(p_T,q_T)$, i.e., 

\[\resizebox{.6\columnwidth}{!}{$\left[\begin{array}{cccccccccccccccccccc}
a &  &  &  &  &  &  &  &  &  &  &  & b &  &  & c & d_T &  &  &  \\
 & a &  &  &  &  &  &  &  &  &  &  &  & b &  &  & c & d_T &  &   \\
 &  & a &  &  &  &  &  &  &  &  &  &  &  & b &  &  & c & d_T &    \\
 &  &  & a &  &  &  &  &  &  &  &  &  &  &  & b &  &  & c & d_T \\
e &  &  & f & g_T &  &  &  &  &  &  &  &  &  &  &  &  &  &  &  \\
 & e &  &  & f & g_T &  &  &  &  &  &  &  &  &  &  &  &  &  &  \\
 &  & e &  &  & f & g_T &  &  &  &  &  &  &  &  &  &  &  &  &  \\
 &  &  & e &  &  & f & g_T &  &  &  &  &  &  &  &  &  &  &  &  \\
 &  &  &  & e &  &  & f & g_T &  &  &  &  &  &  &  &  &  &  &  \\
 &  &  &  &  & e &  &  & f & g_T &  &  &  &  &  &  &  &  &  &   \\
 &  &  &  &  &  & e &  &  & f & g_T &  &  &  &  &  &  &  &  &   \\
 &  &  &  &  &  &  & e &  &  & f & g_T &  &  &  &  &  &  &  &   \\
 &  &  &  &  &  &  &  & e &  &  & f & g_T &  &  &  &  &  &  &   \\
 &  &  &  &  &  &  &  &  & e &  &  & f & g_T &  &  &  &  &  &   \\
 &  &  &  &  &  &  &  &  &  & e &  &  & f & g_T &  &  &  &  &   \\
 &  &  &  &  &  &  &  &  &  &  & e &  &  & f & g_T &  &  &  &   \\
 &  &  &  &  &  &  &  &  &  &  &  & e &  &  & f & g_T &  &  &   \\
 &  &  &  &  &  &  &  &  &  &  &  &  & e &  &  & f & g_T &  &   \\
 &  &  &  &  &  &  &  &  &  &  &  &  &  & e &  &  & f & g_T &   \\
 &  &  &  &  &  &  &  &  &  &  &  &  &  &  & e &  &  & f & g_T
\end{array}\right]$}\;.\]
We computed the resultant $\resultant(p_T,q_T)$ of the two polynomials $p_T$ and $q_T$ with the computer algebra system sage and obtained $\alpha^{40} \beta^{48} \gamma^{32} u^{96} + \alpha^{40}\beta^{48}\gamma^{32}$ as the coefficient of $T^{64}$. Since $\alpha, \beta, \gamma$ are all non-zero, this coefficient is non-zero if and only if $u^{96} \neq 1$. We have $u^{96} = (u^3)^{2^5}$, which is equal to $1$ if and only if $u^3=1$. Since $u$ is not a 7-th power, we have $u^3 \neq 1$, so $\resultant(p_T,q_T)$ is not the zero polynomial.
\end{proof}

\begin{remark}
The upper bound on the linearity of $C_u$ as given in Theorem~\ref{thm:linearity} is tight in the sense that, if $m = 6$ and $u \in \F_{2^m}$ is not a 7-th power, the linearity of $C_u$ matches exactly this bound, i.e., it is equal to $2^{\frac{3m+6}{2}}$. 
\end{remark}

\begin{remark}
Let $m>3$ be odd. Then $8^{1+\lfloor \frac{m}{2} \rfloor}=2^{\frac{3m+3}{2}}$ is the lowest possible linearity that quadratic functions from $(\F_{2^m})^3$ to $(\F_{2^m})^3$ can achieve, unless they are almost bent. Moreover if we can find $(\alpha,\beta,\gamma)\neq(0,0,0)$ such that System~(\ref{eq:linear_structures}) has more than 2 solutions, this would imply that $C_u$ is not almost bent and thus not APN (since a quadratic function is almost bent if and only if it is APN~\cite{DBLP:journals/dcc/CarletCZ98}).  
\end{remark}

\section{Exploring the CCZ-Classes of Quadratic APN Permutations in Dimension 9}
It is not possible at this stage to find the exact number of EA-classes within the CCZ-classes of our permutations, but various techniques allow us to gather a lot of information about the overall structure of each EA-class. Section~\ref{sec:exploring-basics} recalls some necessary concepts, which are then applied in Section~\ref{sec:exploring-overview}. Finally, a similarity between the structures of the CCZ-classes of our functions with the Gold APN permutations is discussed in Section~\ref{sec:gold_similarity}.

\subsection{Tools for Partitioning a CCZ-class}
\label{sec:exploring-basics}
Testing whether two functions are EA-equivalent is difficult from a computational standpoint. However, some EA-class invariants can give us an indirect method to figure out  a lower bound on the number of EA-classes a given set of functions can be partitioned. 

In~\cite{DBLP:journals/ffa/CanteautP19}, it was shown that the thickness spectrum of a function $F\colon \F_2^n \rightarrow \F_2^m$ is an EA-invariant. We also rely on the following quantity.
\begin{definition}[Degree spectrum]
    The \emph{degree spectrum} of $F \colon \F_2^n \rightarrow \F_2^m$ is the multiset $\{d, n_d\}_{d > 0, n_d>0}$ such that $F$ has exactly $n_d$ components of algebraic degree $d$, for all $d$. If $\sum_{d>0}n_d = 2^n-1$ (i.e., if $F$ has no non-trivial affine component), we say that $F$ has a \emph{non-degenerate} degree spectrum.
\end{definition}
\begin{lemma}
    Let $F \colon \F_2^n \rightarrow \F_2^m$ and $G \colon \F_2^n \rightarrow \F_2^m$ be EA-equivalent. If $F$ and $G$ have non-degenerate degree spectra, then said spectra are identical.
\end{lemma}

\begin{proof}
    Let $F : \F_2^n \to \F_2^m$ be a function, and let $C_d$ be the set of its components of algebraic degree exactly $d$, so that the algebraic degree of $x \mapsto \langle c, F(x)\rangle$ is $d$ if and only if $c \in C_d$. By definition, the degree spectrum of $F$ corresponds to the cardinalities of the elements in $\{C_d\}_{d > 0}$. Right-composing $F$ with an affine permutation does not change the algebraic degree of its components, i.e., $x \mapsto \langle c, F(x)\rangle$ and $x \mapsto \langle c, F \circ \eta(x)\rangle$ have the same algebraic degree for all affine permutations $\eta$. Let us now consider left composition. As a constant addition will not change the algebraic degree, we can consider without loss of generality that the composed permutation $\mu$ is linear, i.e. we need to investigate $\eta \circ F$. The components of this function are $x \mapsto \langle c, \eta \circ F(x)\rangle$, which is the same as $x \mapsto \langle \eta^{\top}(c), F(x)\rangle$. Thus, the set of the components of $\eta \circ F$ that have algebraic degree $d$ is exactly $(\eta^{\top})^{-1}(C_d)$, which has the same size as $C_d$.
    
    We now need to consider the addition of an affine function, i.e. to consider $x \mapsto F(x) + \phi(x)$ for some affine function $\phi$. In this case, since it is assumed that no non-trivial component of $F$ or $F + \phi$ is of algebraic degree 1 (or smaller), the addition of $\phi$ cannot modify the algebraic degree of any component.
    
    Overall, provided that no non-trivial component of $F$ is affine, the degree spectrum is constant within an EA-class.
\end{proof}

Using these two spectra, we construct a new form of equivalence between CCZ-equivalent functions. 
\begin{definition}[Region of a CCZ-class]
    Let $F\colon \F_2^n \rightarrow \F_2^m$ and $G \colon \F_2^n \rightarrow \F_2^m$ be two CCZ-equivalent functions. We say that they are in the same \emph{(degree,thickness)-region} (or simply \emph{DT-region}) if they have identical thickness and degree spectra. If the degree spectrum corresponding to a DT-region is non-degenerate, we call the DT-region \emph{non-degenerate}.
\end{definition}
We explicitly mention the (algebraic) degree and the thickness in the name of the DT-region as we hope that yet to be discovered EA-class invariants will eventually allow the definition of even narrower regions.

While a priori cruder than EA-equivalence, this equivalence relation allows to get an already fine-grained view of what is inside the CCZ-class of each of the functions we study. In particular, if two functions are in different non-degenerate DT-regions, they are EA-inequivalent. As a consequence, the number of EA-classes can be bounded as follows.
\begin{lemma}
    Let $z$ be the number of vector spaces of dimension $n$ in the Walsh zeroes of $F \colon \F_2^n \rightarrow \F_2^m$ and let $r$ be the number of non-degenerate DT-regions in its CCZ-class. Then it holds that
    \begin{equation*}
        r ~\leq~ \# \textrm{EA-classes of $F$} ~\leq~ z ~.
    \end{equation*}
\end{lemma}
While the lower bound is easily deduced from the fact that different non-degenerate DT-regions imply different EA-classes, the upper bound was stated in Theorem~4 of~\cite{DBLP:journals/ffa/CanteautP19}.

It was shown in Theorem~3 of~\cite{DBLP:journals/ffa/CanteautP19} that if two functions $F$ and $G$ are CCZ-equivalent but not EA-equivalent, then they both are EA-equivalent to functions $F'$ and $G'$ such that $F'$ and $G'$ are \emph{$t$-twist equivalent}. The $t$ parameter corresponds to the thickness of the vector space contained in $\mathcal{Z}_F$ used to construct the linear permutation applied to $\left\{(x, F(x))\mid x \in \mathcal{V} \right\}$. The specifics of the twisting operation do not matter for our discussion and we refer the interested reader to~\cite{DBLP:journals/ffa/CanteautP19}.

\subsection{An Overview of the CCZ-classes}
\label{sec:exploring-overview}
The CCZ-class of $F_0$ contains 12 non-degenerate DT-regions. Their properties are summarized in Table~\ref{tab:regions-F0}. DT-Region~1 contains $F_0$: as any APN function can only contain one quadratic EA-class (see~\cite{yoshiara2012equivalences}), it is the only non-degenerate DT-region containing quadratic functions. DT-Regions~1 to~6 are those containing permutations, as indicated by the presence of spaces of thickness 9 in the Walsh zeroes of their functions. DT-Region~6 contains the compositional inverse of $F_0$. Overall, for $F_0$ it holds that
\begin{equation*}
    12 ~\leq~ \# \textit{EA-classes of $F_0$} ~\leq~ 4758 ~,
\end{equation*}
and we also remark that permutations in DT-Regions~2 -- 5 are of the shape $(F_0^{-1} + L_i)^{-1}$, where $L_i$ is always a linear function.

\begin{table}
    \centering
    \caption{\normalsize\label{tab:regions-F0}The non-degenerate DT-regions in the CCZ-class of $F_0$. DT-Regions 1 -- 6 contain permutations.}
    \footnotesize
    \begin{tabular}{rcll}
        \toprule
        & twist & deg.\@ spectrum & thickness spectrum \\
        \midrule
        1 & 0 & $\{2: 511\}$ & $\{0: 1, 1: 511, 2: 2590, 3: 1144, 9: 512\}$ \\
        2 & 2 & $\{3: 3, 4: 508\}$ & $\{0: 1, 1: 7, 2: 14, 3: 512, 4: 2576, 5: 1136, 7: 256, 9: 256\}$ \\
        3 & 2 & $\{3: 3, 4: 508\}$ & $\{0: 1, 1: 7, 2: 44, 3: 536, 4: 2546, 5: 1112, 7: 256, 9: 256\}$ \\
        4 & 2 & $\{3: 3, 4: 508\}$ & $\{0: 1, 1: 7, 2: 32, 3: 536, 4: 2558, 5: 1112, 7: 256, 9: 256\}$ \\
        5 & 2 & $\{3: 3, 4: 508\}$ & $\{0: 1, 1: 3, 2: 44, 3: 556, 4: 2546, 5: 1096, 7: 256, 9: 256\}$ \\
        6 & 9 & $\{5: 511\}$ & $\{0: 1, 6: 143, 7: 1295, 8: 2023, 9: 1296\}$ \\ \midrule
        7 & 1 & $\{2: 1, 3: 510\}$ & $\{0: 1, 1: 17, 2: 526, 3: 2574, 4: 1128, 8: 512\}$ \\
        8 & 1 & $\{2: 1, 3: 510\}$ & $\{0: 1, 1: 13, 2: 526, 3: 2578, 4: 1128, 8: 512\}$ \\
        9 & 1 & $\{2: 1, 3: 510\}$ & $\{0: 1, 1: 7, 2: 518, 3: 2584, 4: 1136, 8: 512\}$ \\
        10 & 3 & $\{4: 7, 5: 504\}$ & $\{0: 1, 1: 7, 2: 14, 3: 78, 4: 560, 5: 2506, 6: 1144, 8: 448\}$ \\
        11 & 3 & $\{4: 7, 5: 504\}$ & $\{0: 1, 1: 7, 2: 14, 3: 50, 4: 560, 5: 2534, 6: 1144, 8: 448\}$ \\
        12 & 3 & $\{4: 7, 5: 504\}$ & $\{0: 1, 1: 7, 2: 14, 3: 8, 4: 504, 5: 2576, 6: 1200, 8: 448\}$ \\
        \bottomrule
    \end{tabular}
\end{table}

For $F_1$, there are 19 non-degenerate DT-regions whose properties are listed in Table~\ref{tab:regions-F1}. Those contain functions similar to those in the CCZ-class of $F_1$: DT-Region~1 contains $F_1$, DT-Region~8 contains its inverse, DT-Regions~2 to~7 contain permutations of algebraic degree 4 that are all of the form $(F_1^{-1}+L_i)^{-1}$ for some linear functions $L_i$. The other DT-regions contain functions of algebraic degree 3 and 5 obtained using a 1-twist or a 3-twist (respectively). Interestingly, the functions in DT-Region~15 have a degree spectrum that does not appear in the CCZ-class of $F_0$: while of algebraic degree 5, they have 63 components of algebraic degree 4 instead of only 7. Much like for $F_0$, we have for $F_1$ that
\begin{equation*}
    19 ~\leq~ \# \textit{EA-classes of $F_1$} ~\leq~ 5150 ~.
\end{equation*}

\begin{table}
    \centering
    \footnotesize
    \caption{\normalsize\label{tab:regions-F1}The non-degenerate DT-regions in the CCZ-class of $F_1$. DT-Regions 1 -- 8 contain permutations.}
    \begin{tabular}{rcll}
        \toprule
        & twist & deg.\@ spectrum & thickness spectrum \\
        \midrule
        1 & 0 & $\{2: 511\}$ & $\{0: 1, 1: 511, 2: 2590, 3: 1536, 9: 512\}$ \\
        2 & 2 & $\{3: 3, 4: 508\}$ & $\{0: 1, 1: 7, 2: 56, 3: 512, 4: 2534, 5: 1528, 7: 256, 9: 256\}$ \\
        3 & 2 & $\{3: 3, 4: 508\}$ & $\{0: 1, 1: 7, 2: 44, 3: 560, 4: 2546, 5: 1480, 7: 256, 9: 256\}$ \\
        4 & 2 & $\{3: 3, 4: 508\}$ & $\{0: 1, 1: 7, 2: 38, 3: 536, 4: 2552, 5: 1504, 7: 256, 9: 256\}$ \\
        5 & 2 & $\{3: 3, 4: 508\}$ & $\{0: 1, 1: 7, 2: 38, 3: 560, 4: 2552, 5: 1480, 7: 256, 9: 256\}$ \\
        6 & 2 & $\{3: 3, 4: 508\}$ & $\{0: 1, 1: 3, 2: 46, 3: 556, 4: 2544, 5: 1488, 7: 256, 9: 256\}$ \\
        7 & 2 & $\{3: 3, 4: 508\}$ & $\{0: 1, 1: 7, 2: 50, 3: 560, 4: 2540, 5: 1480, 7: 256, 9: 256\}$ \\
        8 & 9 & $\{5: 511\}$ & $\{0: 1, 6: 192, 7: 1295, 8: 2366, 9: 1296\}$ \\ \midrule
        9 & 1 & $\{2: 1, 3: 510\}$ & $\{0: 1, 1: 21, 2: 518, 3: 2570, 4: 1528, 8: 512\}$ \\
        10 & 1 & $\{2: 1, 3: 510\}$ & $\{0: 1, 1: 17, 2: 534, 3: 2574, 4: 1512, 8: 512\}$ \\
        11 & 1 & $\{2: 1, 3: 510\}$ & $\{0: 1, 1: 19, 2: 534, 3: 2572, 4: 1512, 8: 512\}$ \\
        12 & 1 & $\{2: 1, 3: 510\}$ & $\{0: 1, 1: 15, 2: 534, 3: 2576, 4: 1512, 8: 512\}$ \\
        13 & 1 & $\{2: 1, 3: 510\}$ & $\{0: 1, 1: 15, 2: 526, 3: 2576, 4: 1520, 8: 512\}$ \\
        14 & 3 & $\{4: 7, 5: 504\}$ & $\{0: 1, 1: 7, 2: 14, 3: 106, 4: 504, 5: 2478, 6: 1592, 8: 448\}$ \\
        15 & 3 & $\{4: 63, 5: 448\}$ & $\{0: 1, 1: 3, 2: 14, 3: 94, 4: 616, 5: 2494, 6: 1480, 8: 448\}$ \\
        16 & 3 & $\{4: 7, 5: 504\}$ & $\{0: 1, 1: 7, 2: 14, 3: 78, 4: 616, 5: 2506, 6: 1480, 8: 448\}$ \\
        17 & 3 & $\{4: 7, 5: 504\}$ & $\{0: 1, 1: 7, 2: 14, 3: 92, 4: 616, 5: 2492, 6: 1480, 8: 448\}$ \\
        18 & 3 & $\{4: 7, 5: 504\}$ & $\{0: 1, 1: 7, 2: 14, 3: 64, 4: 616, 5: 2520, 6: 1480, 8: 448\}$ \\
        19 & 3 & $\{4: 7, 5: 504\}$ & $\{0: 1, 1: 7, 2: 14, 3: 64, 4: 560, 5: 2520, 6: 1536, 8: 448\}$ \\
        \bottomrule
    \end{tabular}
\end{table}

\begin{table}
    \centering
    \caption{\normalsize\label{tab:regions-gold}The non-degenerate DT-regions in the CCZ-class of $x \mapsto x^3$ in $\F_{2^9}$. DT-Regions 1 -- 3 contain permutations.}
    \footnotesize
    \begin{tabular}{rcll}
        \toprule
        & twist & deg.\@ spectrum & thickness spectrum \\
        \midrule
        1 & 0 & $\{2: 511\}$ & $\{0: 1, 1: 511, 2: 1022, 3: 584, 9: 512\}$ \\
        2 & 2 & $\{3: 3, 4: 508\}$ & $\{0: 1, 1: 7, 2: 14, 3: 512, 4: 1008, 5: 576, 7: 256, 9: 256\}$ \\
        3 & 9 & $\{5: 511\}$ & $\{0: 1, 6: 73, 7: 511, 8: 1533, 9: 512\}$ \\ \midrule
        4 & 1 & $\{2: 1, 3: 510\}$ & $\{0: 1, 1: 7, 2: 518, 3: 1016, 4: 576, 8: 512\}$ \\
        5 & 3 & $\{4: 7, 5: 504\}$ & $\{0: 1, 1: 7, 2: 14, 3: 8, 4: 504, 5: 1008, 6: 640, 8: 448\}$ \\
        \bottomrule
    \end{tabular}
\end{table}

For comparison, Table~\ref{tab:regions-gold} lists the non-degenerate DT-regions of the Gold APN function $x \mapsto x^3$ in $\F_{2^9}$. DT-Region~1 contains $x \mapsto x^3$, DT-Region~3 contains its inverse, and DT-Region~2 contains a permutation of algebraic degree 4. The other two DT-regions contain functions of algebraic degree 3 and 5 obtained using a 1-twist or a 3-twist (respectively). A similar observation holds for $x \mapsto x^5$ and $x \mapsto x^{17}$, i.e., there are 5 non-degenerate DT-regions in total (having the same degree spectra as DT-Regions 1 -- 5 in Table~\ref{tab:regions-gold}), three of them containing permutations and having the same degree spectra as DT-Regions 1 -- 3 in Table~\ref{tab:regions-gold}. While $x \mapsto x^5$ and $x \mapsto x^{17}$ are not CCZ-equivalent, their CCZ-equivalence classes are partitioned into non-degenerate DT-regions corresponding to identical spectra.

\begin{remark}
We think that the existence of at least 6, resp., 8 pairwise EA-inequivalent permutations within the CCZ-classes of $F_0$, resp., $F_1$ is a quite interesting property and, to the best of our knowledge, the observation of the existence of more than 3 pairwise EA-inequivalent permutations within the CCZ-class of an APN permutation was never reported before. In~\cite{DBLP:journals/ccds/Calderini20}, Calderini computed the number of EA-classes of APN functions in dimension 6 and upper bounds on the number of EA-classes of some APN functions in dimensions 7, 8, and 9. There are exactly two EA-classes containing permutations within the CCZ-class of Dillon's permutation. For all non-Gold APN monomial functions in dimension $n\leq 9$, Calderini observed that their CCZ-classes contain at most two EA-classes. For a Gold APN permutation in odd dimension $n$ divisible by 3, we know from~\cite{DBLP:conf/waifi/Budaghyan07} that its CCZ-class contains at least 3 EA-classes that contain permutations (see also Section~\ref{sec:gold_similarity}), but we are not aware of a Gold APN permutation that contains more than 3 pairwise EA-inequivalent permutations within its CCZ-class. By the same approach based on separating a function into its non-degenerate DT-regions, we verified for $n\leq 9$ that the CCZ-classes of the Gold APN functions contain at most 3 DT-regions that contain permutations. Note that this does not prove that the number of pairwise EA-inequivalent permutations within the CCZ-class of such a function is bounded above by 3. 
\end{remark}

\subsection{A Similarity of $F_0$ and $F_1$ to Gold APN Permutations}
\label{sec:gold_similarity}
Budaghyan showed in~\cite{DBLP:conf/waifi/Budaghyan07} that for a Gold APN permutation $G_i \colon \F_{2^n} \rightarrow \F_{2^n}, x \mapsto x^{2^i+1}$ in odd dimension $n$ divisible by three, we can obtain a permutation of algebraic degree 4 as $(G_i^{-1}+L_i)^{-1}$, where $L_i\colon \F_2^n \rightarrow \F_2^n, x \mapsto \trace_{n,3}(x+x^{2^{2i}})$. Here, $\trace_{n,3}\colon x \mapsto x + x^{2^3} + x^{2^{2\cdot3}} + \dots + x^{2^{n-3}}$ denotes the relative trace function from $\F_{2^n}$ into the subfield $\F_{2^3}$.  This is an example that, in general, it is possible to obtain an APN permutation EA-inequivalent to a permutation $F$ and its inverse by just applying EA-transformations and inversion to $F$.

We will now see that a similar property also holds for $F_0$ and $F_1$, more precisely we show this for the EA-equivalent permutations $C_u$. For a permutation $F \colon (\F_{2^m})^t \rightarrow (\F_{2^m})^t$ and a linear mapping $L \colon (\F_{2^m})^t \rightarrow (\F_{2^m})^t$, we define the mapping $T_{F,L} \colon x \mapsto F^{-1}(x) +L(x)$. 

\begin{theorem}
\label{prop:degree_4_general}
Let $t \in \mathbb{N}, t\geq 2$ and let $F \colon (\F_{2^3})^t \rightarrow (\F_{2^3})^t$ be a permutation of the form
\[ \left(\begin{array}{c} x_1 \\ \vdots \\ x_i \\ \vdots \\ x_t \end{array}\right)^{\top} \mapsto \left(\begin{array}{c} x_1^3 + f_1(x_2,x_3,\dots,x_{t-1},x_t)  \\ \vdots \\ x_i^3 + f_i(x_1,\dots,x_{i-1},x_{i+1}, \dots,x_t) \\ \vdots \\ x_t^3 + f_t(x_1,x_2,\dots,x_{t-2},x_{t-1}) \end{array}\right)^{\top}, \quad x_i \in \F_{2^3}\;,\]
where $f_i \colon (\F_{2^3})^{t-1} \rightarrow \F_{2^3}$, and let \[L \colon (\F_{2^3})^t \rightarrow (\F_{2^3})^t, (x_1,x_2,\dots,x_t) \mapsto (x_1 + x_1^4,0,0, \dots,0)\;.\] Then, $T_{F,L}$ is a permutation. If we further have that, for $2 \leq i \leq t$, the functions $f_i$ as functions in $(x_1,\dots,x_{t-1})$ are of algebraic degree at most 2 and such that each  monomial $x_1^{2^i+2^j}, i\neq j$ vanish and if $f_1$ is of algebraic degree at most 2, then $T_{F,L}^{-1}$ is of algebraic degree at most 4. If further $f_1^6 + f_1^5 + f_1^3$ has algebraic degree 4, we have that the algebraic degree of $T_{F,L}^{-1}$ is equal to 4. 
\end{theorem}
\begin{proof}
For the function $P \coloneqq T_{F,L} \circ F$, we have
\begin{align*} P\left(\begin{array}{c} x_1 \\ x_2 \\ \vdots \\ x_t \end{array}\right)^{\top} &= \left(\begin{array}{c} x_1 \\ x_2 \\ \vdots \\ x_t \end{array}\right)^{\top} + L\left(F \left( \begin{array}{c} x_1 \\ x_2 \\ \vdots \\ x_t \end{array} \right)^{\top} \right)^{\top} = \left( \begin{array}{c} x_1^5 + x_1^3 + x_1 + h(x_2,\dots,x_t) \\ x_2 \\ \vdots \\ x_t \end{array}\right)^{\top},\end{align*}
where $h(x_2,\dots,x_t) \coloneqq f_1(x_2,\dots,x_t) + f_1^4(x_2,\dots,x_t)$. Since $X^5 + X^3 + X \in \F_{2^3}[X]$ (i.e., the Dickson polynomial of degree 5) is a permutation polynomial of $\F_{2^3}$, the function $P$ is a permutation. Then, $T_{F,L}$ is a permutation because $F$ is a permutation.

Let us now proceed by proving the statement on the algebraic degree of $T_{F,L}^{-1}$. The inverse of $p \colon \F_{2^3} \rightarrow \F_{2^3}, x \mapsto x^5 + x^3 + x$ can be given as $q \colon \F_{2^3} \rightarrow \F_{2^3}, x \mapsto x^5 + x^4 + x^3 + x^2 + x$, which is of algebraic degree $2$. We then have
\begin{align*} T_{F,L}^{-1}\left(\begin{array}{c} x_1 \\ x_2 \\ \vdots \\ x_i \\ \vdots \\ x_t \end{array}\right)^{\top} &= F \circ P^{-1}\left(\begin{array}{c} x_1 \\ x_2 \\ \vdots \\ x_i \\ \vdots \\ x_t \end{array}\right)^{\top} = F \left(\begin{array}{c} q(x_1+h(x_2,\dots,x_t)) \\ x_2 \\ \vdots \\ x_i \\ \vdots \\ x_t \end{array}\right)^{\top} \\ &= \left(\begin{array}{c} q(x_1+h(x_2,\dots,x_t))^3 + f_1(x_2,\dots,x_t) \\ x_2^3 + f_2(y,x_3,x_4,\dots,x_t)  \\ \vdots \\ x_i^3 + f_i(y,x_2, x_3,\dots, x_{i-1},x_{i+1}, \dots,x_t) \\ \vdots \\ x_t^3 + f_t(y,x_2,x_3,\dots,x_{t-1})  \end{array}\right)^{\top},\end{align*}
where $y \coloneqq q(x_1+h(x_2,\dots,x_t))$. Since both $q$ and $h$ are functions of algebraic degree at most 2, the function $(x_1,x_2,\dots,x_t) \mapsto y$ is of algebraic degree at most 4. Moreover, since $q^3(x) = x^3 + x^2 +x$ is also of algebraic degree 2, we further have that $(x_1,x_2,\dots,x_t) \mapsto q(x_1+h(x_2,\dots,x_t))^3$ is of algebraic degree at most 4. For deducing that the algebraic degree of $T_{F,L}^{-1}$ is bounded above by 4, it is left to show that all the functions
\[ \tilde{f}_i \colon (x_1,x_2,\dots,x_t) \mapsto f_i(y,x_2,x_3,\dots,x_{i-1},x_{i+1},\dots,x_t), \quad 2 \leq i \leq t\]
are of algebraic degree at most 4. Because of our assumptions, each $\tilde{f}_i$ for $2 \leq i \leq t$ can only consist of monomials of the form $y^{2^{a}}x_j^{2^{b}}$ or $x_j^{2^{a}}x_k^{2^{b}}$ with $a,b \in \mathbb{N}$, $2 \leq j \leq k \leq t$. The monomials not involving $y$ are of algebraic degree at most 2, so it is enough to focus on those monomials involving $y$. By denoting $\bar{h} \coloneqq h(x_2,\dots,x_t)$, we have
\[ y = q(x_1+\bar{h}) = x_1^5 + x_1^4 + x_1^3 + x_1^2 + x_1 + x_1^4\bar{h} + x_1^2\bar{h} + x_1\bar{h}^4 + x_1\bar{h}^2 + \bar{h}^4 + \bar{h}^2 +\bar{h} + \bar{h}^3 + \bar{h}^5\] and it is left to show that $\bar{h}^3 + \bar{h}^5$ only consist of monomials of algebraic degree at most 3.
Indeed, we have
\begin{align*}
    \bar{h}^3 + \bar{h}^5 &= \left(f_1(x_2,\dots,x_t) + f_1(x_2,\dots,x_t)^4\right)^3 + \left(f_1(x_2,\dots,x_t) + f_1(x_2,\dots,x_t)^4\right)^5 \\
    &= f_1(x_2,\dots,x_t) + f_1(x_2,\dots,x_t)^2\;,
\end{align*}
which only consist of monomials of algebraic degree at most 2. Thus, the algebraic degree of $T_{F,L}^{-1}$ is bounded above by 4. To show that the algebraic degree of $T_{F,L}^{-1}$ is at least 4, we observe that
\[ q^3(x_1 + \bar{h}) = x_1^3 + x_1^2 + x_1 + x_1^2\bar{h} + x_1\bar{h}^2 + \bar{h} + \bar{h}^2 + \bar{h}^3\;,\]
which consists of a monomial of algebraic degree 4 if and only if $\bar{h}^3$ consists of a monomial or algebraic degree 4. Since 
\[ \bar{h}^3 = f_1(x_2,\dots,x_t)^6 + f_1(x_2,\dots,x_t)^5 + f_1(x_2,\dots,x_t)^3 + f_1(x_2,\dots,x_t)^2\;,\]
this is the case if $f_1^6 + f_1^5 + f_1^3$ is of algebraic degree 4.
\end{proof}

\begin{corollary}
Let $u\in \F_{2^3} \setminus \{0,1\}$ and let $C_u \colon (\F_{2^3})^3 \rightarrow (\F_{2^3})^3, \quad (x,y,z) \mapsto (x^3+uy^2z, y^3+uxz^2,z^3+ux^2y)$. Let $L \colon (\F_{2^3})^3 \rightarrow (\F_{2^3})^3, (x,y,z) \mapsto (x + x^4,0,0)$. Then $T_{C_u,L}^{-1}$ is an APN permutation of algebraic degree 4.
\end{corollary}
\begin{proof}
The function $C_u$ with $u \in \F_{2^3} \setminus \{0,1\}$ is a permutation of the form as in Theorem~\ref{prop:degree_4_general} with $t = 3$ and $f \coloneqq f_1 = f_2 = f_3\colon (x,y) \mapsto u x^2 y$. Since $f$ is of algebraic degree 2, contains no monomial of the form $x^3, x^5$, or $x^6$ and since $(f^6 + f^5 +f^3)(x,y) = u^3y^6z^3 + u^5y^3z^5 + u^6y^5z^6$ contains monomials of algebraic degree 4, we have that the algebraic degree of $T_{C_u,L}^{-1}$ is equal to 4. The APN-ness of $T_{C_u,L}^{-1}$ follows since $C_u$ is APN and $T_{C_u,L}^{-1}$ is by construction CCZ-equivalent to $C_u$.
\end{proof}

In particular, taking $u \in \F_{2^3}$ as a zero of $X^3+X+1$, we obtain a permutation 
\[ T_{C_u,L}^{-1} = \left( C_u^{-1} + \left( x+x^4, 0, 0 \right) \right)^{-1}\]
that lies in DT-Region 2 of Table~\ref{tab:regions-F0}. Taking $u \in \F_{2^3}$ as a zero of $X^3+X^2+1$, we obtain a permutation $T_{C_u,L}^{-1}$ that lies in DT-Region 2 of Table~\ref{tab:regions-F1}. 

We will see now that the condition $m=3$ is necessary in order to obtain permutations $T_{C_u,L}$ similar as those stated in Theorem~\ref{prop:degree_4_general}. We first show a more general statement.

\begin{theorem}
\label{prop:general_noperm}
Let $m,k,t \in \mathbb{N}$, $m$ be odd, $t\geq 2$, $\gcd(k,m)=1$ and let $F \colon (\F_{2^m})^t \rightarrow (\F_{2^m})^t$ be a permutation of the form
\[ \left(\begin{array}{c} x_1 \\ \vdots \\ x_i \\ \vdots \\ x_t \end{array}\right)^{\top} \mapsto \left(\begin{array}{c} x_1^{2^k+1} + f_1(x_2,x_3,\dots,x_{t-1},x_t)  \\ \vdots \\ x_i^{2^k+1} + f_i(x_1,\dots,x_{i-1},x_{i+1}, \dots,x_t) \\ \vdots \\ x_t^{2^k+1} + f_t(x_1,x_2,\dots,x_{t-2},x_{t-1}) \end{array}\right)^{\top}, \quad x_i \in \F_{2^m}\;,\]
where $f_i \colon (\F_{2^m})^{t-1} \rightarrow \F_{2^m}$, and let \[L \colon (\F_{2^m})^t \rightarrow (\F_{2^m})^t, (x_1,x_2,\dots,x_t) \mapsto (x_1 + x_1^{2^{2k}},0,0, \dots,0)\;.\] Then, $T_{F,L}$ is a permutation if and only if $m=3$. 
\end{theorem}
\begin{proof}
By the same argument as in the first part of the proof of Theorem~\ref{prop:degree_4_general}, the function $T_{F,L}$ is a permutation if and only if the polynomial $X^{(2^k+1) \cdot 2^{2k}} + X^{2^k+1} + X$ is a permutation polynomial of $\F_{2^m}$. For $m=3$, we only need to consider the cases of $k=1$ and $k=2$, so the polynomials $X^{12} + X^3 + 1$ and $X^{80} + X^5 + X$, which both correspond to the permutation polynomial $X^5 + X^3 + X$ in $\F_{2^3}$. Since $n$ is odd, we have $\gcd(2k,m) = \gcd(k,m) = 1$. Then, the statement for $m>3$ follows from Lemma~\ref{lem:permpoly} below.
\end{proof}

\begin{lemma}
\label{lem:permpoly}
Let $n>3$ be odd and let $i,j \in \mathbb{N}$ be such that $\gcd(i,n) = \gcd(j,n)=1$. Then $X^{(2^i+1)\cdot 2^j} + X^{2^i+1} + X$ is not a permutation polynomial of $\F_{2^n}$. 
\end{lemma}
\begin{proof}
From $n>3$, we deduce the existence of an element $\alpha \in \F_{2^n} \setminus \{0\}$ with $\trace(\alpha) = 0$ and $\trace(\alpha^{-(2^i+1)}) = 1$. To do so, we first deduce the existence of an element $x \in \F_{2^n} \setminus \{0\}$ with $\trace(x) \neq \trace(x^{-(2^i+1)})$. Indeed, suppose that we have $\trace(x) = \trace(x^{-(2^i+1)})$ for all elements $x \in \F_{2^n} \setminus \{0\}$, the function \[G \colon x \mapsto \trace(x^{2^n-2^i-2}) = \begin{cases}
    \trace(x^{-(2^i+1)}) &\text{if } x \neq 0 \\
    0 &\text{if } x=0 
\end{cases} \] would be identical to $x \mapsto \trace(x)$, which is linear. However, this can only happen for $n \leq 3$.  Let us now partition the set $\F_{2^n}$ into the disjoint union of the four sets $S_{k,l}$ for $k, l \in \F_2$, where
\[S_{k,l} \coloneqq \{x \in \F_{2^n} \mid \trace(x) = k, \trace(x^{2^n-2^i-2}) = l\}\;.\]
We have $|S_{0,0}|+|S_{0,1}| = |S_{1,0}|+|S_{1,1}| = |S_{0,0}|+|S_{1,0}| = |S_{0,1}|+|S_{1,1}| = 2^{n-1}$, since $G$ is a permutation and thus each of these unions corresponds exactly to either $\trace(x)$ or $\trace(x^{2^n-2^i-2})$ being a constant. We deduce from this chain of equality that $|S_{0,1}| = |S_{1,0}| = 2^{n-1} - |S_{0,0}|$. Since we established that $|S_{1,0}|+|S_{0,1}| > 0$, we deduce that both $|S_{1,0}| > 0$ and $|S_{0,1}| > 0$.

Let us therefore fix an element $\alpha \in \F_{2^n} \setminus \{0\}$ with $\trace(\alpha) = 0$ and $\trace(\alpha^{-(2^i+1)}) = 1$. We show that the equation
\begin{equation}
\label{eq:permpoly}
    x^{(2^i+1)\cdot 2^j} + x^{2^i+1} + x + (x + \alpha)^{(2^i+1) \cdot 2^j} + (x+\alpha)^{2^i+1} + (x+\alpha) = 0
\end{equation}
has a solution $x \in \F_{2^n}$. By substituting $x$ with $\alpha x$ in Equation~(\ref{eq:permpoly}), we obtain
\begin{equation} \label{eq:permpoly_eq}\left(\alpha^{2^i+1} (x^{2^i}+x+1)\right)^{2^j} + \alpha^{2^i+1} (x^{2^i} + x + 1) = \alpha\;.\end{equation}
Since $\gcd(j,n)=1$, we have that $\gcd(2^j-1,2^n-1) = 1$, so the kernel of $x \mapsto x^{2^j} + x$ is exactly $\{0,1\}$. Then, the image of $x \mapsto x^{2^j} + x$ is the set $\{ \beta \in \F_{2^n} \mid \trace(\beta) = 0\}$, and we can write $\alpha = \beta^{2^j} + \beta$ for some element $\beta \in \F_{2^n} \setminus \{0,1\}$.
Since $\gcd(i,n)=1$ and $\trace(1)=1$, the image of $x \mapsto x^{2^i}+x+1$ is exactly $\{ z \in \F_{2^n} \mid \trace(z) = 1\}$, so Equation~(\ref{eq:permpoly_eq}) (and thus also Equation~(\ref{eq:permpoly})) has a solution if there exists a $z \in \F_{2^n}$ with $\trace(z) = 1$ such that $(\alpha^{2^i+1} z)^{2^j} + \alpha^{2^i+1} z = \alpha$, or equivalently, 
\begin{equation}\label{eq:permpoly2}(\alpha^{2^i+1} z + \beta)^{2^j} + (\alpha^{2^i+1}z+\beta) = 0\;.\end{equation}
Since the kernel of $x \mapsto x^{2^j} + x$ is $\{0,1\}$, the solutions  $z$ of Equation~(\ref{eq:permpoly2}) are exactly those that fulfill $z = \beta \alpha^{-(2^i+1)}$ or $z = \beta \alpha^{-(2^i+1)} + \alpha^{-(2^i+1)}$. We deduce that such a solution $z$ with $\trace(z)=1$ exists since $\trace(\alpha^{-(2^i+1)}) = 1$.
\end{proof}

This yields the following corollary for the case of $C_u$.
\begin{corollary}
Let $m \in \mathbb{N}$ and let $C_u \colon (\F_{2^m})^3 \rightarrow (\F_{2^m})^3, (x,y,z) \mapsto (x^3+uy^2z, y^3+uxz^2,z^3+ux^2y)$ be a permutation. Then, $C_u^{-1}+(x+x^4,0,0)$ is a permutation if and only if $m=3$.
\end{corollary}

\section{Conclusion and Open Problems}
As a first approach to generalize the recently-found APN permutations $F_0$ and $F_1$ in dimension 9 into infinite families, we derived a trivariate representation $C_u$ of those APN permutations over $(\F_{2^m})^3$ for $m=3$ and analyzed the differential uniformity and linearity in the general case of $m>3$. We also analyzed the CCZ-classes of $F_0$ and $F_1$ in more detail and observed that they contain many EA-classes containing permutations. Similar as for Gold APN permutations in odd dimension $n$ divisible by $3$, it is possible to derive a permutation EA-inequivalent to both $F_0$ (resp., $F_1$) and its inverse by just applying EA transformation and inversion to $F_0$ (resp., $F_1$). 

As open problems, it would be interesting to prove the second part of Conjecture~\ref{con:1}, i.e., prove that $C_u$ with $u \in \F_{2^m} \setminus \{0\}$ not being a 7-th power is not a permutation for $m>3$, and to investigate in which cases $C_u$ is CCZ-equivalent to a permutation. For dimension 9 particularly, it would be interesting to determine the exact number of EA-classes containing permutations within the CCZ-classes of $F_0$ and $F_1$, not just a lower bound, and to further analyze how those EA-inequivalent permutations can be obtained by just applying EA-transformation and inversion to $F_0$ and $F_1$. Still, the most interesting open question is whether $F_0$ and $F_1$ can be generalized into an infinite family of APN permutations.

\subsection*{Acknowledgment}
We thank the anonymous reviewers for their valuable comments and suggestions to improve the quality of the paper.
We further thank Lilya Budaghyan for some useful discussion at an early stage of this project.

\end{document}